\documentclass[twocolumn]{aastex61}
\pdfoutput=1 
\usepackage{amsmath,amstext}
\usepackage[T1]{fontenc}
\usepackage{apjfonts} 
\usepackage[figure,figure*]{hypcap}

\usepackage{natbib}

\newcommand{\eg}{\textit{e.g.}, }
\newcommand{\ie}{\textit{i.e.}, }

\newcommand{\TW}{\textwidth}
\newcommand{\CW}{\columnwidth}
\newlength{\imsize}
\shorttitle{17 February 2013 sunquake}
\shortauthors{Green et al.}

\begin{document}

\title{The 17 February 2013 sunquake in the context of the active region's magnetic field configuration}

\author[0000-0002-0053-4876]{Green, L.M.}
\author[0000-0001-7809-0067]{Valori, G.}
\affiliation{Mullard Space Science Laboratory, UCL, Holmbury St. Mary, Dorking, Surrey, RH5 6NT, UK}
\author{Zuccarello, F. P.}
\affiliation{Centre for Mathematical Plasma Astrophysics, Department of Mathematics, KU Leuven, Celestijnenlaan 200B, B-3001 Leuven, Belgium}
\affiliation{Mullard Space Science Laboratory, UCL, Holmbury St. Mary, Dorking, Surrey, RH5 6NT, UK}
\author[0000-0002-2348-3522]{Zharkov, S.}
\affiliation{Department of Physics and Mathematics,
University of Hull, Hull, HU6 7RX, UK}
\author[0000-0001-9346-8179]{Matthews, S.A.}
\affiliation{Mullard Space Science Laboratory, UCL, Holmbury St. Mary, Dorking, Surrey, RH5 6NT, UK}
\author[0000-0002-1837-2262]{Guglielmino, S.L.}
\affiliation{Dipartimento di Fisica e Astronomia - Sezione Astrofisica, Universit\`{a} degli Studi di Catania, Via S. Sofia 78, 95123 Catania, Italy}

   \date{Received September 15, 1996; accepted March 16, 1997}

 
\begin{abstract}

Sunquakes are created by the hydrodynamic response of the lower atmosphere to a sudden deposition of energy and momentum. In this study we investigate a sunquake that occurred in NOAA active region 11675 on 17 February 2013. Observations of the corona, chromosphere and photosphere are brought together for the first time with a non-linear force-free model of the active region's magnetic field in order to probe the magnetic environment in which the sunquake was initiated. We find that the sunquake was associated with the destabilization of a flux rope and an associated M-class GOES flare. Active region 11675 was in its emergence phase at the time of the sunquake and photospheric motions caused by the emergence heavily modified the flux rope and its associated quasi-separatrix layers, eventually triggering the flux rope's instability. The flux rope was surrounded by an extended envelope of field lines rooted in a small area at the approximate position of the sunquake. We argue that the configuration of the envelope, by interacting with the expanding flux rope, created a "magnetic lens" that may have focussed energy in one particular location the photosphere, creating the necessary conditions for the initiation of the sunquake.

\end{abstract}


%

\section{Introduction}

A sunquake is a sub-photospheric pressure perturbation (acoustic wave) that propagates into the solar interior, where the increasing plasma temperature and sound speed cause the wave to be refracted back to the photosphere. When the acoustic wave reaches the photosphere, the sharp gradient in the plasma conditions reflects the wave back into the Sun. This process of refraction and reflection creates circular wavefronts in the photosphere around the site of the sunquake origin. The existence of sunquakes was first suggested by \cite{Wolff72} who proposed that acoustic perturbations could be created as a consequence of the energy conversion and magnetic restructuring that takes place during a solar flare. This hypothesis was finally supported when observations taken with the MDI instrument onboard SOHO captured images of the characteristic circular wavefronts of a sunquake in Doppler velocity data \citep{KosovichevZharkova1989}. This discovery has naturally led to investigations aimed at understanding the mechanisms that deliver the momentum impulse to the solar interior in order to drive a sunquake. 

Observations have shown that all sunquakes occur in concert with a solar flare. Using a wide variety of datasets, the point of initiation of the sunquake has been seen to be co-spatial with hard X-ray emission \citep{Donealindsey05,Kosovichev2007},
white light emission \citep{Doneaetal2006,Buitrago-Casasetal15},
gamma ray emission \citep{ZharkovaZharkov2007}
and abrupt and permanent changes in the photospheric line-of-sight magnetic field \citep{KosovichevZharkova01}. These observations are in line with the standard model for solar flare formation, which involves the acceleration of non-thermal particles from the corona down to the lower atmosphere, where they deposit their energy and momentum and heat the photospheric/chromospheric plasma. In light of this, these observations are often linked to sunquake initiation mechanisms in the following ways. 

Hard X-ray and gamma ray emission support the idea of sunquakes being produced by hydrodynamic shocks created by sudden thick target heating of the upper and middle chromosphere either by electron \citep{KosovichevZharkova1989,Kosovichev2007} or proton beams \citep{ZharkovaZharkov2007}. 
White light emission indicates that heating of the photosphere is occurring and this observation led to the proposal of the backwarming mechanism for sunquake formation \citep{Doneaetal2006}. In this scenario, particle beams heat the chromosphere and chromospheric radiation, at Balmer and Paschen wavelengths, heats the photosphere. 
The observed magnetic field changes led to the suggestion that a Lorentz force or McClymont jerk \citep{Hudsonetal08,Fisheretal12} might be responsible for sunquake generation since a sharp variation in the Lorentz force could create a transient pressure wave.
More recently plasma heating by waves has been a proposed as a possible mechanism for sunquake formation \citep{RussellFletcher13,Matthewsetal15}. 


A complementary approach to studying the electromagnetic radiation signatures is to probe the configuration of the magnetic field at the time of the sunquake. Free magnetic energy is thought to be the source of energy used to power a sunquake (sunquakes typically have an energy in the range $10^{27}$ to $10^{29}$ ergs). Furthermore, the magnetic field configuration acts as a guide for non-thermal particles and waves, influencing the sites of momentum and energy deposition that could be related to sunquake formation. The above listed sunquake formation mechanisms viewed in the context of the standard flare model normally invoke (or indeed observe) the location of sunquake initiation to be at the footpoints of flare loops. However, observations of strong hard X-ray, white light emission or magnetic field changes in the feet of flare loops are no guarantee that a sunquake will be produced. Sunquakes have also been observed away from the flare loops (and hence from the main location of energy deposition by non-thermal particles) at the photospheric footpoints of an erupting flux rope \citep{Zharkovetal11}. It is therefore essential to investigate the coronal magnetic structure in order to understand the origin of sunquakes.

In this study we investigate where in the magnetic configuration of NOAA active region 11675 the sunquake of 17 February 2013 occurred, to seek an understanding of how the magnetic field  influences the regions of the photosphere where energy and momentum are deposited.
Our approach uses observations alongside a non-linear force-free magnetic field extrapolation. In Section \ref{s:sunquake} we discuss the sunquake detection and characteristics, in Section \ref{s:observations} the observations, Section \ref{s:nlfff} the non-linear force-free field extrapolation and in Sections \ref{s:discussion} and \ref{s:conclusion} we discuss our findings and make our conclusions.


\section{The 17 February 2013 sunquake}
\label{s:sunquake}

Full disk Solar Dynamics Observatory (SDO) Helioseismic and Magnetic Imager (HMI) observations \citep{Scherreretal12} are used to detect and investigate the sunquake that occurred on 17 February 2013 in NOAA active region 11675. 45 second cadence HMI data are selected over an 8-hour time window. The data are prepared in the following way. The region of interest around the active region is extracted, the data are then remapped using the Postel projection technique and the differential rotation is removed using the Snodgrass differential rotation rate. The spatial resolution of the remapped data is 0.04 heliographic degrees per pixel. The heliographic centre of the extracted region is 37E,12N at 11:59 UT on 17 February 2013, and the data series is extracted on 17 February 2013 from 11:59 UT to  19:58 UT. 

\begin{figure}
\centering
\includegraphics[width=0.5\textwidth]{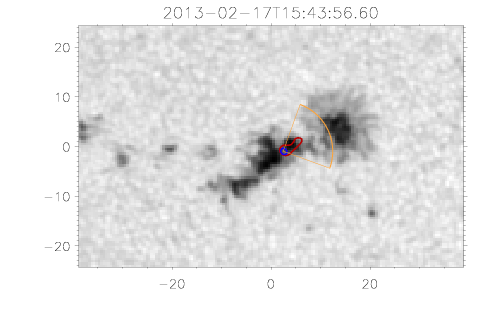}\\
\includegraphics[width=0.5\textwidth]{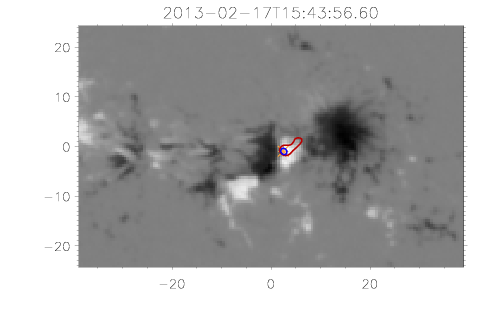}
\caption{\label{fig:egression_ic_mg} Top panel: 5~mHz (blue) and 6~mHz (red) egression power contours overlaid on the co-temporal Postel remapped HMI continuum image. The orange arc indicates the direction in which the acoustic emission was most intense. Bottom panel: (blue) and 6 (red) mHz egression power contours overlaid on the co-temporal Postel remapped HMI line-of-sight magnetic field image. Here and in the following maps, north is up and west to the right.}
\end{figure}

The acoustic holography technique described in 
\cite{Zharkovetal11,Zharkovetal13}
is used to create "egression" power maps from the HMI velocity observations. 
The egression power is computed for each integral frequency from 3 to 10~mHz by applying a 1~mHz frequency bandwidth filter to the data. Green's functions, built for a surface monochromatic point source of the corresponding frequency using a geometrical optics approach, are used. The computation essentially backtracks the observed acoustic disturbances to a point on the photosphere from which they originate, hence revealing the likely sunquake initiation region. The egression power maps show a well-defined and statistically significant (at the 5-sigma level) kernel in the 5-10~mHz bands indicating the occurrence of a sunquake, which provides an indication of the source region. Figure~\ref{fig:egression_ic_mg} 
show the 5~mHz (blue) and 6~mHz (red) egression power contours (at factor 4 times the mean quiet Sun egression power) overplotted on the Postel remapped intensity and line-of-sight HMI data. It can be seen that the egression kernel is cospatial with a region of positive polarity magnetic field in NOAA active region 11675.

The 6mHz kernel (red contour in Figure~\ref{fig:egression_ic_mg}) is elongated in the north-west direction, possibly indicating a moving source. The egression emission takes place from 15:42:26 to 15:57:26 UT on 17 February, however, the choice of a 1 mHz bandwidth filtering in the pre-processing induces the following small timing uncertainty in the egression power data

\[\Delta t = \frac{1}{\Delta \nu} = \frac{1}{0.001} seconds \] 

\begin{figure}
\centering
\includegraphics[width=0.5\textwidth]{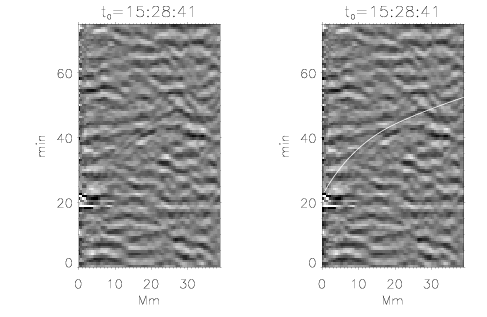}
\caption{\label{fig:full_td_diagram}Full time-distance diagram showing the ridge associated with the outward propagating acoustic source. The panel on the right hand-side indicates the best fit curve to the ridge (white line). Both panels use the same time range on the y-axis with zero corresponding to the time 17 February 2013 15:28:41 UT.} 
\end{figure}

In order to determine the time of the sunquake more precisely, and verify and improve the location determination, we  also compute directional and full time-distance diagrams (see Figure \ref{fig:full_td_diagram} for the full time-distance diagram)
from the running difference Postel-remapped and tracked dopplergrams \citep[for a comparison of the two techniques see][]{Zharkovetal13}. 
The location of the acoustic source as determined from the time-distance diagram is indicated by an orange asterisk in the bottom panel of Figure~\ref{fig:egression_ic_mg}.
The full time-distance diagram (where the data are integrated over a whole circle rather than a chosen arc) shows a noticeable ridge that is present at the obtained location ($\pm$ 1-2 pixels in each direction). The ridge is more pronounced in the directional diagram, with the direction indicated in Figure~\ref{fig:egression_ic_mg}, top panel,
by an orange arc emanating from the source. The directional holography analysis confirms that most of the acoustic emission indeed went in that direction. To obtain the time of the quake, we fit the ridge seen in the time-distance diagram with the theoretical acoustic travel-time, showing the sunquake start time to be 15:50:41-15:51:26 UT.  This is later than the time given by \cite{Sharykinetal15}, where the sunquake initiation time is estimated based on a photospheric velocity transient at the location of their source. Our source is close to that of \cite{Sharykinetal15}, a little over 2Mm to the South East, and from 15:46:56 UT it is also affected by transient photospheric velocity changes. The results of our ridge fitting give a sunquake timing near the end of the velocity transient.

There are three main sources of timing uncertainty, associated with using the time-distance technique, that we take into account. First, any data related issues (e.g. cadence of the data). In our case, 45 second cadence running difference HMI Doppler data are used, which leads to a 90 second timing uncertainty (since running difference data are Image(n+1) - Image(n)). Second, there is a source of uncertainty related to the visual ridge identification, ridge properties and curve fitting. To investigate this uncertainty we analysed three time-distance curves corresponding to different initiation times. Using $t_{0}$ in Figure \ref{fig:full_td_diagram} as a reference, these times are $t_{0}+22$ minutes, $t_{0}+21$ min and $t_{0}+23$ minutes. Considering these results we find a good agreement with our original timing estimate. The third source of timing uncertainty has to do with the angular width of the ripples (and the time-distance diagram ridge) and the accuracy of the source location. When computing a directional time-distance diagram over a relative small arc (90 degrees or smaller), it is possible to move the time-distance diagram around the original location with recomputed diagrams often also showing a ridge that may have different properties and start time. To investigate this uncertainty we examine time-distance ridges from a number of origins to determine the best possible sunquake location.


\begin{figure}
\centering
\includegraphics[width=0.4\textwidth]{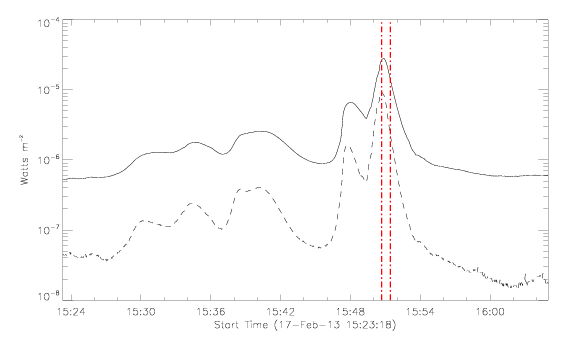}
\caption{GOES light curve with the quake time interval (including uncertainty) indicated between the vertical red dash-dot lines.}
 \label{fig:goes_lightcurve}
\end{figure}

The time of the sunquake as determined from the directional time-distance analysis shows that the sunquake occurred in NOAA active region 11675 close in time to two flares: a C-class flare and an M-class flare. The timing of the C-class and M-class flares can be seen in Figure~\ref{fig:goes_lightcurve}, with the sunquake time range indicated between the vertical red dashed lines. The sunquake occurs around the time that the soft X-ray emission from the M-class flare is seen to peak. 

The sunquake energy has been calculated following the method of \cite{Zharkovetal13}. In order to determine the sunquake energy, the acoustic egression is computed applying a 1-mHz bandwidth filter so that the measurements at each frequency band do not overlap. A snapshot of the 6-mHz egression power is then used to define the sunquake source area. The egression power is determined by integrating over a box that encompasses the source area. 
Varying the spatial box and the time-window of the integration allows us to compute the mean acoustic energy as 1.68$\times 10^{28}$ erg, with an error given via standard deviation estimated at $\pm0.34\times10^{28}$ ergs.
The sunquake energy can be compared to the typical energy of the non-thermal electrons associated with a M-class flare, which has been found to be of the order of $10^{30}$ erg \citep{SaintHilaire05}.

\section{Observations}
\label{s:observations}

In this section we discuss the evolution of NOAA active region 11675 before, during and after the occurrence of the sunquake. In addition to HMI, data from SDO's Atmospheric Imaging Assembly \citep{Lemen2012}, GOES X-ray sensors, SOHO's Large Angle Spectrometric Coronagraph \citep{Brueckner95}, STEREO's COR 1 coronagraph \citep{Thompson2003} and RHESSI \citep{lin2002} are also used.

\subsection{Photospheric magnetic field evolution}
\label{s:photosphere}

The evolution of the photospheric magnetic field is studied using line-of-sight and vector magnetic field data from the SDO/HMI instrument. Data at a cadence of 45s (which have a noise level of 10.2 Gauss) and 720s (which have a noise level of 6.3 Gauss) are used \citep{Liu2012a}. Both HMI datasets have a pixel size of 0.5\arcsec.

The formation of NOAA active region 11675 is observed to take place on the Earth-facing side of the Sun. In this section we describe the evolution of the active region from its first appearance to the day of the sunquake. Overall, the active region evolution is dominated by the emergence of a series of bipoles with associated motions, collision of polarities and, in one bipole, the splitting of its trailing polarity. Collectively, this evolution creates a magnetic configuration that includes a polarity inversion line along which the magnetic field is highly sheared. This polarity inversion line becomes the focus of our study in later sections. 

\begin{figure*}
   \centering
   \includegraphics[width=0.9\textwidth]{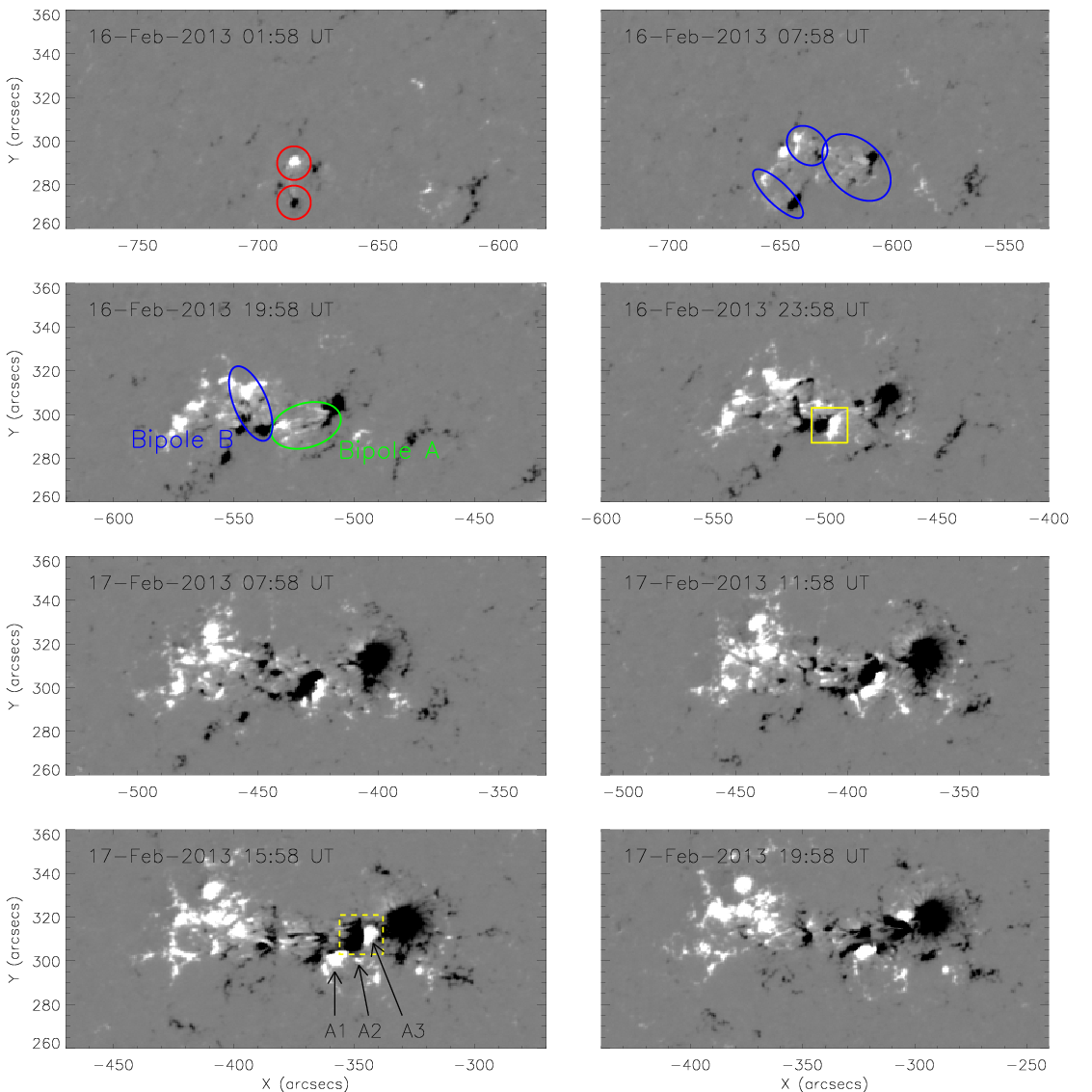}
      \caption{The formation and evolution of NOAA active region 11675 as seen in HMI line of sight data saturated between $\pm$500 Gauss. The red circles in the top left panel indicate the first bipole emergence. Successive bipole emergence is indicated by blue and green ellipses. The yellow solid-line box indicates the polarity inversion line formed by the collision of Bipole A and Bipole B. The yellow dashed-line box in the final panel indicates the polarity inversion line along which the activity studied originates.}
         \label{fig:LOS_evolution}
   \end{figure*}

The initial emergence of the flux of NOAA active region 11675 is characterised by a bipole that begins to appear at the end of 15 February 2013, and which has a north-south field orientation (indicated by the red circles in top-left panel of Figure~\ref{fig:LOS_evolution}). Three more bipoles emerge near-simultaneously. These bipoles are visible by 16 February 2013 around 05:30 UT. All three bipoles have negative leading polarity and are inclined to the north-south direction by roughly 45 degrees in the clockwise direction (indicated by the blue ellipses in top-right panel of Figure~\ref{fig:LOS_evolution}). Flux cancellation takes place between the two northern-most bipoles, where there is a collision of opposite polarity magnetic field.

A second stronger episode of flux emergence begins around 16 February 2013 16:30 UT. This time the magnetic field emerges into an already complex region, and the emergence is again composed of several bipoles; there is the emergence of a bipole in the east of the active region which has a negative leading polarity (and which is inclined by around 45 degree to the east-west line), a bipole in the central part of the active region that has a north-south orientation, and a bipole which has its leading negative polarity emerging in the western side of the active region. The location of the latter bipole is significant and is indicated by the green ellipse in Figure~\ref{fig:LOS_evolution}. We call this bipole "Bipole A". The negative flux of Bipole A merges with the pre-existing negative field on the western side of the active region and the positive polarity collides with the negative flux of the north-south oriented emerging bipole to the east (we call this "Bipole B"). 

Overall, the negative polarities of the bipoles emerging in this second phase migrate westwards to join the developing leading negative polarity sunspot. Because of the motions and the ongoing emergence there is a region where the positive flux of Bipole A collides with the negative flux of Bipole B (indicated by the yellow box in Figure~\ref{fig:LOS_evolution}). After the collision, the positive flux of Bipole A fragments and splits into three parts, indicated in the bottom left panel of Figure~\ref{fig:LOS_evolution} by A1, A2 and A3. The northern most fragment, A3, exhibits a clear motion between 17 February 11:00 UT and 17 February 16:00 UT, when it moves rapidly north-westwards by approximately~14\arcsec~(10 Mm). Giving the positive polarity fragment a plane-of-the-sky speed of approximately 0.5 kms$^{-1}$. The sunquake is located in this positive polarity fragment (A3), very close to the polarity inversion line. The negative polarity of Bipole B is also moving north-west as it "seeks out" the active region's negative polarity leading spot. The coalescence of like-polarity field is a common feature of active region formation \citep{vandrielgesztelyigreen15}. In the following section we discuss the evolution of the corona and how it is influenced by the evolution of the photospheric magnetic field. 

The HMI cylindrical equal area (CEA) vector data allow a study of the magnetic field structure at the photosphere. The data reveal that there are locations where the magnetic field vector is tangential to the photosphere. These regions are known as bald patches \citep{titov93}. The locations of the bald patches can be seen in the left panel of Figure \ref{fig:BP}. The bald patches are present in particular along the PIL between the fragment A3 and the negative field of Bipole B. The bald patches are relatively stable and persistent between 17 February 12:00 UT and 17 February 15:48 UT (the time when the C-class flare emission peaks). After this time bald patches along this section of the PIL are no longer observed. The field lines associated with the bald patches are described in Section~\ref{s:aia_comparison}.

\subsection{Evolution of the corona}
\label{s:corona}

\begin{figure}
    \centering
    \includegraphics[width=\columnwidth]{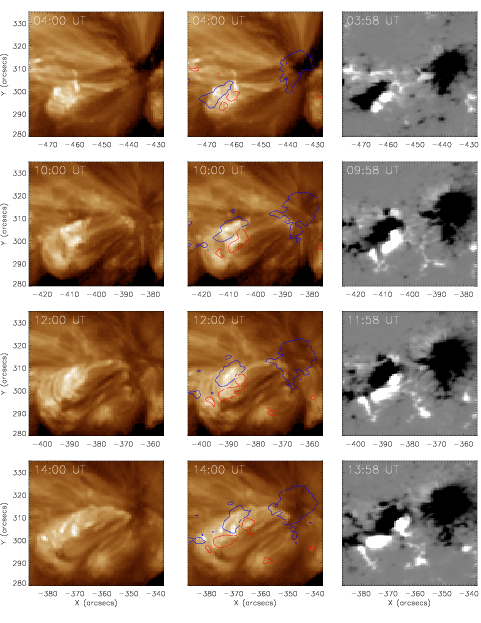}
       \caption{The evolution of the corona on 17 February 2013 as seen in the 193~\AA{} channel (left and middle columns) along with the co-aligned HMI line-of-sight magnetic field data (right column). The middle column shows the contours of the  positive field (red) and the negative field (blue) at the 500G level overlaid on the AIA data.}
       \label{fig:corona_evolution_193}
    \end{figure}

We provide here an overview of the evolution of the coronal field that is driven by the flux emergence and associated photospheric motions that are discussed in Section \ref{s:photosphere}. On a global scale, the active region magnetic field appears to not be highly sheared. However, at low altitudes the field appears to be more non-potential. When the positive polarity fragment of Bipole A (indicated in the green ellipse in Figure \ref{fig:LOS_evolution}) emerges its corresponding negative field is located to the west. Magnetic connections therefore exist between these two polarities. However, the collision between the positive flux of Bipole A and the negative flux of Bipole B to the east creates magnetic connections between these polarities too. The splitting of the positive polarity of Bipole A, and the strong northwest motions of fragment A3 along with negative polarity of Bipole B, shear the coronal field. Figure \ref{fig:corona_evolution_193} shows these motions of the photospheric magnetic field and the AIA emission structures associated with the coronal magnetic field configuration, indicating the magnetic connectivities.

\begin{figure}
\centering
\includegraphics[width=0.4\textwidth]{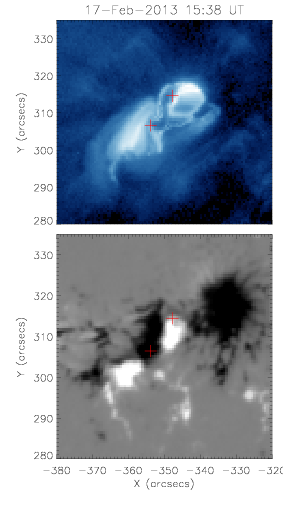}
\caption{AIA 335~\AA{} waveband image (top panel) showing the location of the small filament. Red crosses indicate the approximate end points of the filament and show that it lies along the polarity inversion line of the sheared negative and positive polarities, and above the sunquake location (bottom panel).}
\label{fig:AIA_335_HMI_1538}
\end{figure}

The AIA data also show the formation of a small filament along the polarity inversion line in between A3 and  negative polarity of Bipole B. This filament first becomes visible in the AIA wavebands around 15:24 UT (although filament material may have existed before this time) and it is more visible in the AIA 335 ~\AA{} waveband by 17 February 2013 15:31 UT. The southern end of the filament then darkens and becomes slightly larger so that the filament is well observed by 17 February at 15:38 UT when the filament has a slight forward S shape to it (top panel of Figure~\ref{fig:AIA_335_HMI_1538}).

\begin{figure*}
\centering
\includegraphics[width=0.8\textwidth]{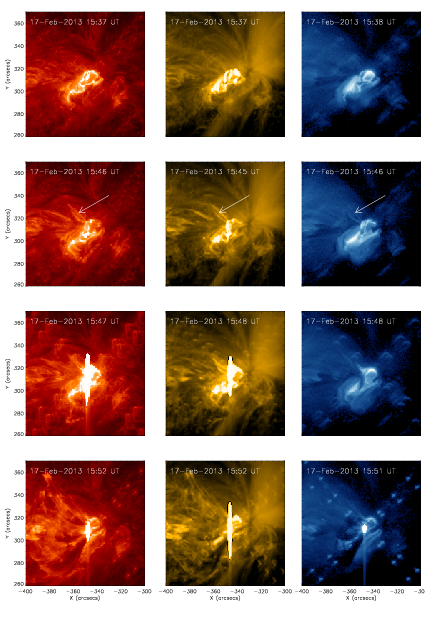}
\caption{AIA images in the 304~\AA{}, 171~\AA{} and 335~\AA{} wavebands showing the evolution of the corona above the sunquake origin. The arrows in the panels of the second row point to an initial linear brightening that appears to be connected to the event. Later, this activation spreads to the south and becomes a broad fan of emitting and absorbing material, associated with a filament eruption, that can be seen in the across all wavebands in the bottom panels.
}
\label{fig:aia_images}
\end{figure*}

After 15:38 UT the filament becomes harder to observe, presumably either due to plasma heating, which would cause filament plasma to go from absorption to emission, or by the emission of plasma higher in altitude than the filament material. By 15:46 UT some thin brightened threads above the northern end of the filament have appeared. In the plane of the image these brightened threads extend off to the east. These brightened threads can be seen in the second row of Figure~\ref{fig:aia_images} and are indicated by arrows.

\subsection{C and M class flaring activity}\label{s:flare}

The activity discussed in this section includes a C-class flare that began around 15:46 UT (peak ~15:48 UT) and an M-class around 15:49 UT (peak ~15:51 UT). A careful study of the timeline of events is needed due to the very close temporal coincidence of these two flares.

The filament (shown in Figure \ref{fig:AIA_335_HMI_1538}) is still in place at the start of the C-class flare but by the peak of this flare (15:48 UT) absorbing filament material is seen to be being ejected upwards along field lines that run parallel to the pre-flare brightened threads. Absorbing filament material continues to flow upwards during the M-class flare and is observed along field lines further to the south as the event proceeds (bottom row of Figure~\ref{fig:aia_images}). LASCO C2 and C3 coronagraph data do not detect a CME associated to either the C-class or M-class flare. There was however, an EUV wave that appeared to propagate into the corona and a narrow and faint mass outflow was visible in STEREO COR1-B data \citep{Rom2014}. These observations indicate that the filament erupted but did not fully escape the corona. Therefore, we could class this as a failed eruption.

\begin{figure*}
\centering
\includegraphics[width=0.9\textwidth]{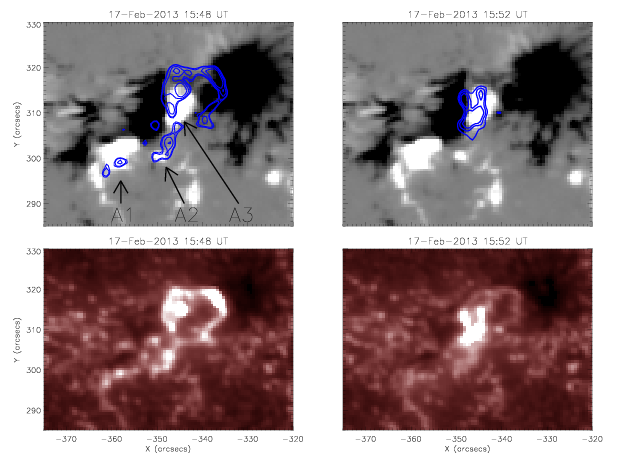}
\caption{AIA 1700~\AA{} images showing the ribbons at the peak of the C-class flare (left column) and the M-class flare (right column). The HMI images (top row) are aligned with the AIA 1700 images (bottom row) and show the flare ribbon locations as blue contours. In both images, the circular ribbon is present, which is associated with a fan structure found in the NLFFF model. The circular ribbon is formed during the C-class flare (this flare perturbs the null point structure) and the emission is fading during the M-class event. HMI data are saturated between $\pm$500 Gauss.}
\label{fig:1700_ribbons}
\end{figure*}

The flare ribbons seen in the AIA 1700~\AA~channel (Figure~\ref{fig:1700_ribbons}) indicate that these flares were indeed two separate energy releases rather than a single event that exhibited a complex soft X-ray profile. This is evidenced by the observations in the 1700~\AA~waveband which show that the flare ribbons are in slightly different locations, and have a different morphology, for the two flares. The ribbons associated to the C-class flare exhibit a circular-shaped emission that extends along the negative polarity sunspot and round into the field of the positive polarity fragment A3 of Bipole A (left-hand column of Figure~\ref{fig:1700_ribbons}). There is also a bright spot-like emission in the northern section of the positive polarity. The circular shape of a ribbon was demonstrated to be generally associated with the presence at coronal heights of a null point, \cite[see \textit{e.g.}][and references therein]{Masson2017}, with the circular ribbon marking the location where the fan associated with the null intersects the chromosphere.
This is in contrast to the M-class flare, during which the emission appears over almost the entirety of the positive polarity fragment A3 and over the
the polarity inversion line that exists between the positive polarity of polarity A3 and the negative polarity of Bipole B (right-hand column of Figure~\ref{fig:1700_ribbons}). The ribbons of the M-class flare are more reminiscent of the configuration of a two-ribbon flare arcade that is produced in the standard model configuration of an erupting flux rope/sheared core field.

The sunquake is closely coupled in time with these two flares, but with the sunquake initiation time calculated as being in the time range 15:50:41 to 15:51:26 UT ($\pm$ 90 seconds), the sunquake is more likely initiated during the impulsive phase of the M-class flare rather than the C-class flare (which peaked at 15:48 UT). It is worth noting that in the thick target model for sunquake generation by the deposition of energy into the lower atmosphere by accelerated electrons, there is expected to be a delay of round 100 to 200 seconds between the particles giving up their energy and the hydrodynamic response of the plasma that initiates the sunquake. The hard X-ray energy estimates, as determined from the RHESSI data are 6.6$\times10^{26}$ erg$s^-1$ at the peak of the M-class flare and 1.5$\times10^{28}$ erg$s^-1$ at the peak of the C-class flare. It is noted that the C-class flare shows a more impulsive hard X-ray lightcurve than the M-class. 
This may intuitively imply that the sunquake would therefore be more likely to be associated with the C-class flare. However, the hard X-ray energy estimates reflect the non-thermal energy produced during the flare, rather than the total energy. The penetration depth of the non-thermal electrons influences the ratio of thermal to non-thermal emission as there is a depth in the atmosphere at which the energy transported by more deeply penetrating electrons is radiated away and does not drive chromospheric evaporation \citep[see for example,][]{mcDonald99}.


\section{Non-linear force-free field modelling}
\label{s:nlfff}
In order to probe the magnetic field configuration in the corona and investigate the likely sites of energy release, a non-linear force-free field (NLFFF) extrapolation is computed using a HMI vector magneotgram taken on 17 February 2013 at 14:36 UT. This time was chosen because it represents a snapshot of the active region prior to the sunquake, when no dynamic activity is seen taking place. The rationale behind the use of a NLFFF model at the time selected is that the global field of the active region is not expected to be significantly evolving during the 1h and 20m between the time of the extrapolation and the sunquake. The main change in the line-of-sight magnetogram data visible during this interval is the northward migration of the positive polarity of Bipole A (fragment A3) that shears the field even further, and which may drive magnetic reconnection. Indeed small brightenings are observed in the AIA 1700~\AA{} channel over A3 in the time between the NLFFF extrapolation and the sunquake. These small brightenings spread to the negative polarity of bipole B and become more intense around 15:20 UT. However these brightenings are not expected to change the global structure of the active region.

Since the photospheric magnetic field is not force-free, the vector magnetogram was "preprocessed" in order for it to be used as the boundary condition for the force-free field extrapolation. The preprocessing method used is described in \cite{Fuhrmann2007}. In the application used here, only the transverse component of the field was modified by the preprocessing, with a maximum variation in each pixel equal to the largest of 100 G and 30\% of the local value. These maximal ranges of variation resulted in an average modification of 76~G (respectively, 81~G) in the $B_{\rm x}$ (respectively, $B_{\rm y}$) component, and in a decrease of the total Lorentz force on the magnetogram from 0.24 before preprocessing to 0.08 after preprocessing, according to the forcing definition used in \cite{Metcalf2008}.
Hence, the preprocessed magnetogram retains relatively strong forcing, despite preprocessing. This is likely due to the relatively large distance from disk center of the active region (about 22$^\circ$ East, 19$^\circ$ North, which is about half the solar radius), where the noise is relatively large \citep[see, \eg Fig.3 in][]{Liu2012a} and the less accurate, and arguably more noisy, transverse component enters the vertical component in a larger proportion.

The preprocessed vector magnetogram was then extrapolated with the method discussed in \cite{valori2010} using three levels of grid refinement, to obtain a model of the coronal field. As a consequence of the residual, non-negligible forces on the magnetogram, the resulting extrapolated coronal model is only approximately force- and divergence-free. The fraction of the current that is perpendicular to the field in the volume is around 0.5, with the largest contribution in the height between $z=$3 and $z=$20. Correspondingly, the errors due to the violation of the solenoidal property are also not negligible, amounting to 9.5\% of the total magnetic energy according to the estimation from \cite{Valori2013}.
Due to these limitations of the extrapolation, it is important to verify to what extent the  NLFFF is a representative model of the coronal magnetic field, by comparing it with AIA observations at 14:36 UT.

\subsection{NLFFF extrapolation and its comparison to AIA data}
\label{s:aia_comparison}

\begin{figure*}
\setlength{\imsize}{0.5\TW}
\includegraphics[width=\textwidth]{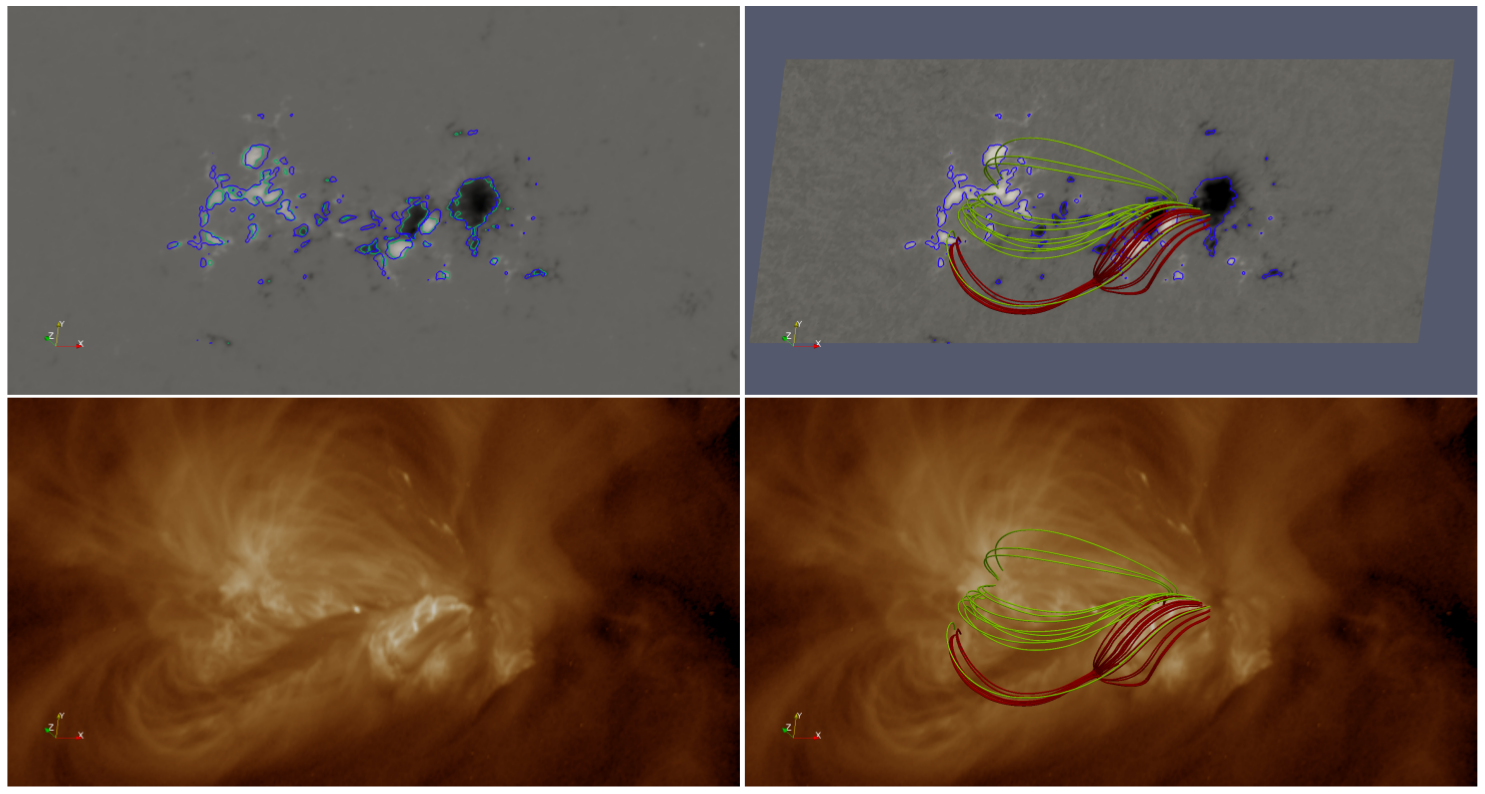}
\caption{\label{fig:AIA_NLFFF_global_comparison} 
\textbf{Top row:} 
HMI line-of-sight magnetogram in greyscale (saturated at $\pm$1500~G) and a 500~G green isoline of the field amplitude (\textbf{left}). The purple line is the 500~G isoline of the amplitude of the line-of-sight field computed from the vector magnetogram. The difference between the green and the purple isolines is a measure of the best attainable alignment. On the \textbf{right}, field lines starting at the coronal null point (red) defining the spine and fan separatices, and selected large-scale field lines (green) started right outside the (photospheric QSL print corresponding to the) fan, see also Figure~\ref{fig:mgm_qsl}. The vertical component of the vector magnetogram (greyscale, saturated at $\pm$1500~G) is plotted in the background.   
\textbf{Bottom row:} 
AIA 193~\AA{} without (\textbf{left}) and with (\textbf{right}) the same selected field lines representing some of the visible large-scale field. The time of the AIA images in this figure is at the same as the vector magnetogram employed for the extrapolation, \ie 14:36 UT 17 February 2013. All panels are in the image plane, that is, from the observer viewpoint.
}
\end{figure*}

\begin{figure*}
\setlength{\imsize}{0.5\TW}
\includegraphics[width=\textwidth]{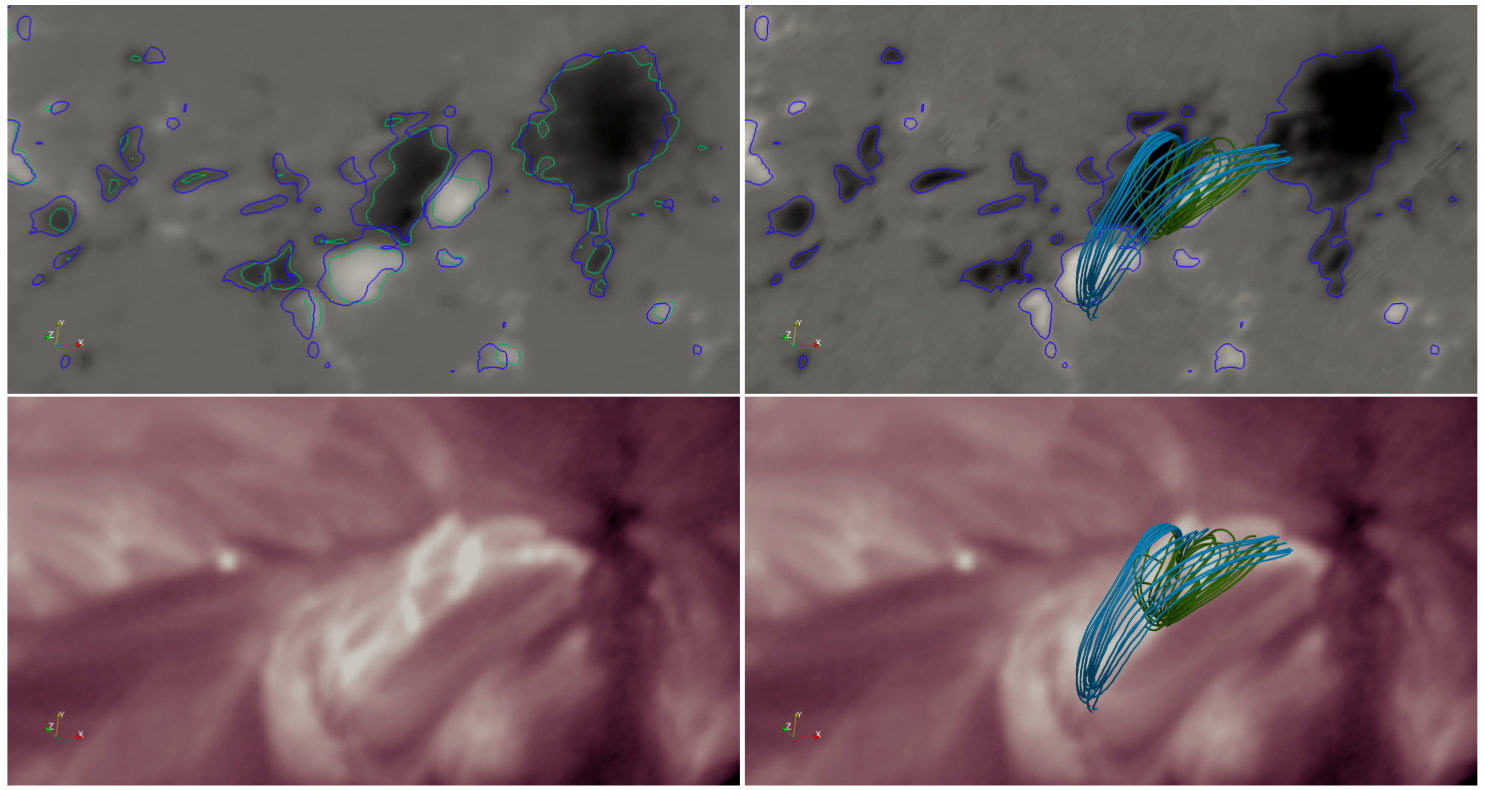}
\caption{\label{fig:AIA_NLFFF_core_comparison} 
\textbf{Top row:} 
HMI line-of-sight magnetogram in greyscale (saturated at $\pm$1500~G) and a 500~G green isoline of the field amplitude (\textbf{left}). The purple line is the 500~G isoline of the amplitude of the line-of-sight field computed from the vector magnetogram.
On the \textbf{right}, two groups of selected large-scale field lines (cyan and green) overplotted onto the vertical component of the vector magnetogram (greyscale, saturated at $\pm$1500~G). The lower group of (green) field lines is referred to in the following as ``envelope field lines'', cf. Figure~\ref{fig:HFT}.
\textbf{Bottom row:} 
AIA 211~\AA{} without (\textbf{left}) and with (\textbf{right}) the same selected field lines representing some of the visible large-scale field.
The time of the AIA images in this figure is the same as that of the vector magnetogram employed for the extrapolation, \ie 14:36 UT of 17 February 2013.
All panels are in the image plane, that is from the observer viewpoint.
}
\end{figure*}

To assess the global validity of the NLFFF model, we verify that AIA features can be reliably identified as emitting/absorbing plasma on field lines bundles that have a correspondence to selected field lines in the extrapolation. For such a comparison, AIA and HMI images need to be co-aligned with the extrapolation. 
AIA images and line-of-sight magnetograms (obtained from the Doppler camera) are plane-of-sky images, whereas the vector magnetogram data (obtained by inverting the filtergrams) have been transformed using the CEA projection method. The relatively large distance from disk center of the active region  enhances the differences between the two projections (and instruments). A compromise between these different views is obtained by matching isocontours of magnetic field from the two instruments in the central area of the vector magnetogram. In particular, the plane-of-sky and extrapolation co-alignment is obtained by scaling the line-of-sight magnetogram until the isocontours of the line-of-sight magnetogram match the isocontours of the line-of-sight field reconstructed from the vector magnetogram data (viewed as from the SDO satellite, \ie in parallel projection on the image plane). The result of the comparison is presented in the top left panels of Figure~\ref{fig:AIA_NLFFF_global_comparison} and  Figure~\ref{fig:AIA_NLFFF_core_comparison}.  The differences between the instruments recording the data, as well as their subsequent manipulations, show clearly in the  mismatch between the (green and purple) isolines of the two images. Such a mismatch is considered to be a lower limit to the accuracy with which we can match NLFFF extrapolation and AIA images, since the NLFFF extrapolation cannot possibly attain a higher level of matching than the input data it is based on.

Once the two line-of-sight magnetograms have been co-aligned, the geometrical transformation obtained for the line-of-sight image is applied to the AIA images. Field lines from the extrapolation can then be compared to the transformed AIA images. Note that, depending on the selected filter, the formation altitude of the emitting structures seen in AIA can vary considerably, especially considering the off-disk-center position of the active region. Therefore, such a procedure is necessarily approximate. 

The field lines in the right-hand column of Figure \ref{fig:AIA_NLFFF_global_comparison} show a selection of large-scale features of the extrapolated field compared with the 193~\AA{} image at the time of the extrapolation. An additional confirmation of the quality of the extrapolation at large scales is the presence of a null with its associated fan/spine structure. Such a configuration is supported by the observational data through the shape of the flare ribbons shown in Figure \ref{fig:1700_ribbons} and discussed in Section \ref{s:flare}. A null point is indeed found in the model of the coronal field. The red field lines in Figure~\ref{fig:AIA_NLFFF_global_comparison} start from the null region, and delineate the spine and fan separatrices. There is a very good match between the footprint of the fan and the circular ribbon position, despite the time difference between the extrapolation and the flare, as will be shown in more detail in Section \ref{s:discussion}. Moreover, the green field lines in Figure~\ref{fig:AIA_NLFFF_global_comparison} that start right outside the photospheric print of the fan (cf. also Figure~\ref{fig:mgm_qsl} and Section \ref{s:qsl}) qualitatively match some of the emission strutures seen in the AIA~193~\AA{} image. These field lines are also involved in the dynamic phase of the eruption accompanying the flare activity, see Section \ref{s:discussion}. 

At a smaller scale, Figure~\ref{fig:AIA_NLFFF_core_comparison} shows a comparison between the observed sheared field above the polarity inversion line between the positive polarity of Bipole A and the negative polarity of Bipole B and the extrapolated field. Field lines that have been noted in the pre-flare evolution of the photospheric and coronal field, discussed in Section \ref{s:corona} in relation to Figure~\ref{fig:corona_evolution_193}, are also found in the extrapolation. Figure \ref{fig:AIA_NLFFF_core_comparison} shows in cyan and green lines remnants of the original connectivity of the positive and negative field of Bipole A as well as the new connections made between the positive field of Bipole A and the negative field of Bipole B.
These cyan and green field lines belong to two different domains of connectivity (see the QSL analysis below for more details). Figure \ref{fig:AIA_NLFFF_core_comparison} shows that the NLFFF extrapolation includes sets of field lines that, combined, qualitatively correspond to the emission structures observed in AIA~211\AA{} at the time of the extrapolation. In conclusion, we are confident that the NLFFF extrapolation captures the essential topological elements of the magnetic configuration both at large and small scales despite the limitations of the extrapolation discussed above.

At the time of the extrapolation, field lines associated with the bald patches reported in Section \ref{s:photosphere} can be investigated. The bald patch field lines are actually very short, confirming the observations reported in Section \ref{s:photosphere}. The middle and right-hand panels of Figure~\ref{fig:BP} show field lines all starting along a vertical line located at the central bald patch position. Three groups of field lines are depicted in red, orange and yellow which change concavity and connectivity in a relatively short span in altitude. These field lines are compatible with the presence of a hyperbolic flux tube (HFT) in the coronal field, separating the red from the orange field lines. See Section \ref{s:qsl} for more details.

\setlength{\imsize}{0.33\TW}
\begin{figure*}
\begin{center}
\includegraphics[width=\imsize]{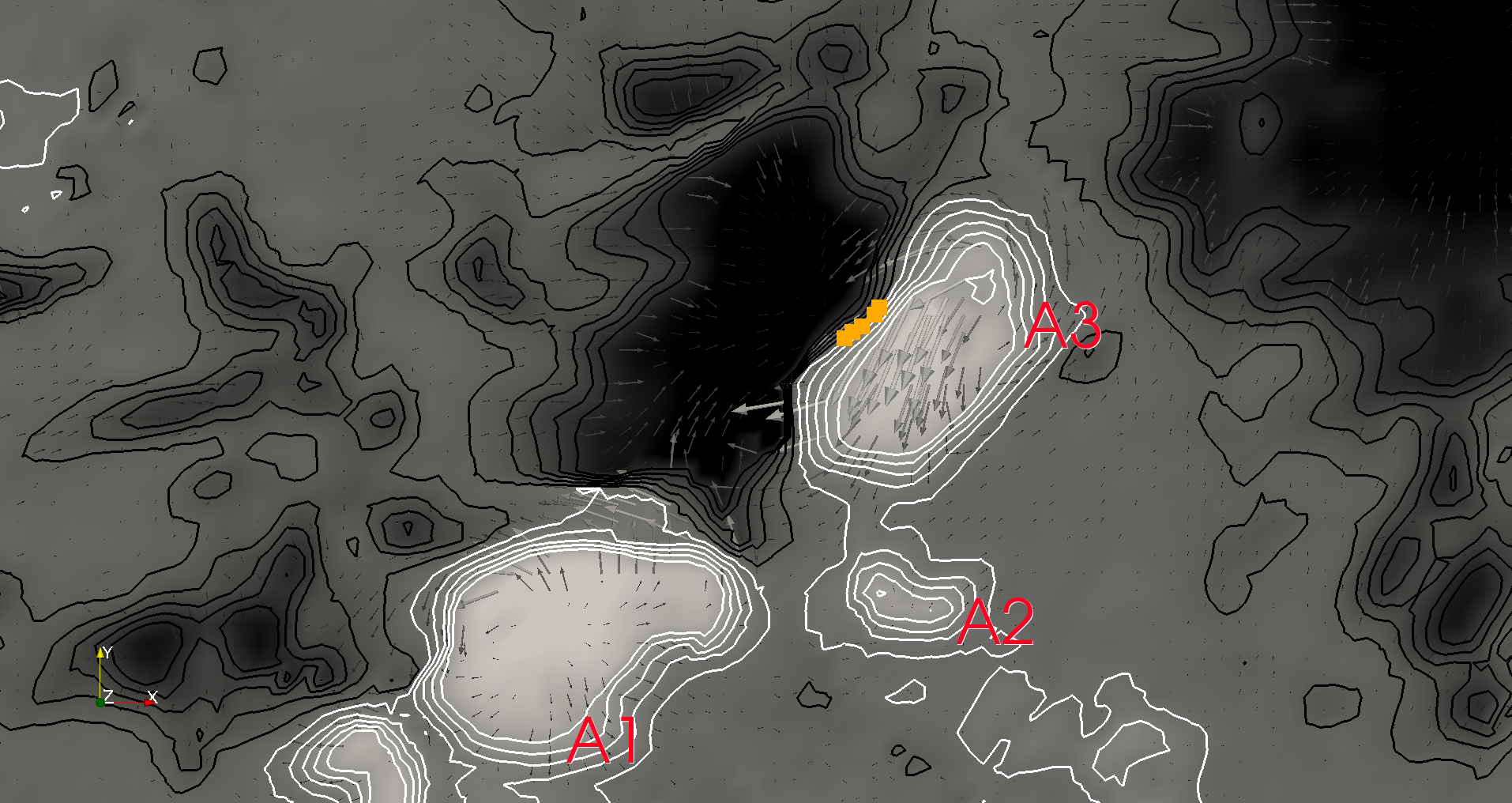}
\includegraphics[width=\imsize]{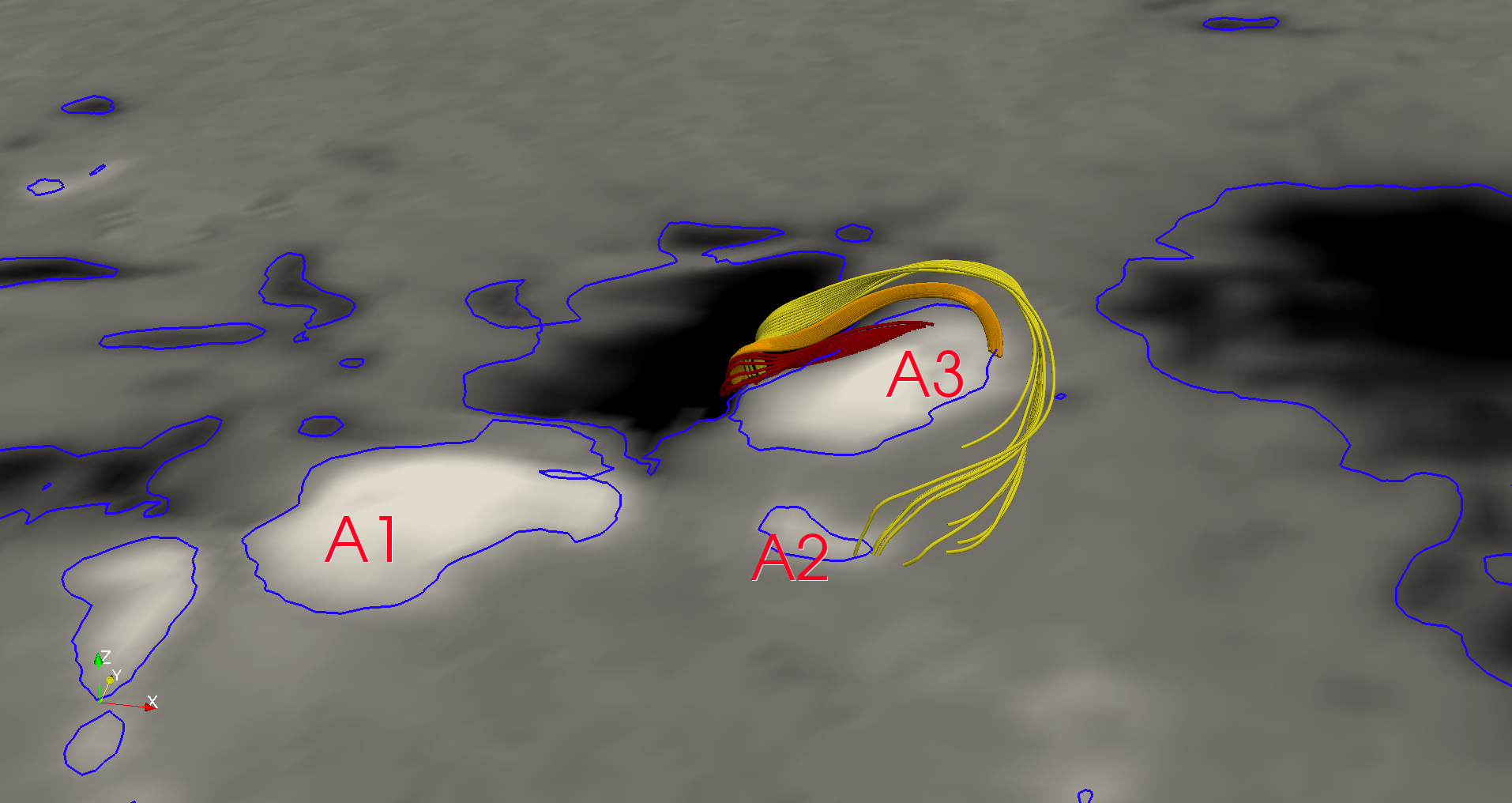}
\includegraphics[width=\imsize]{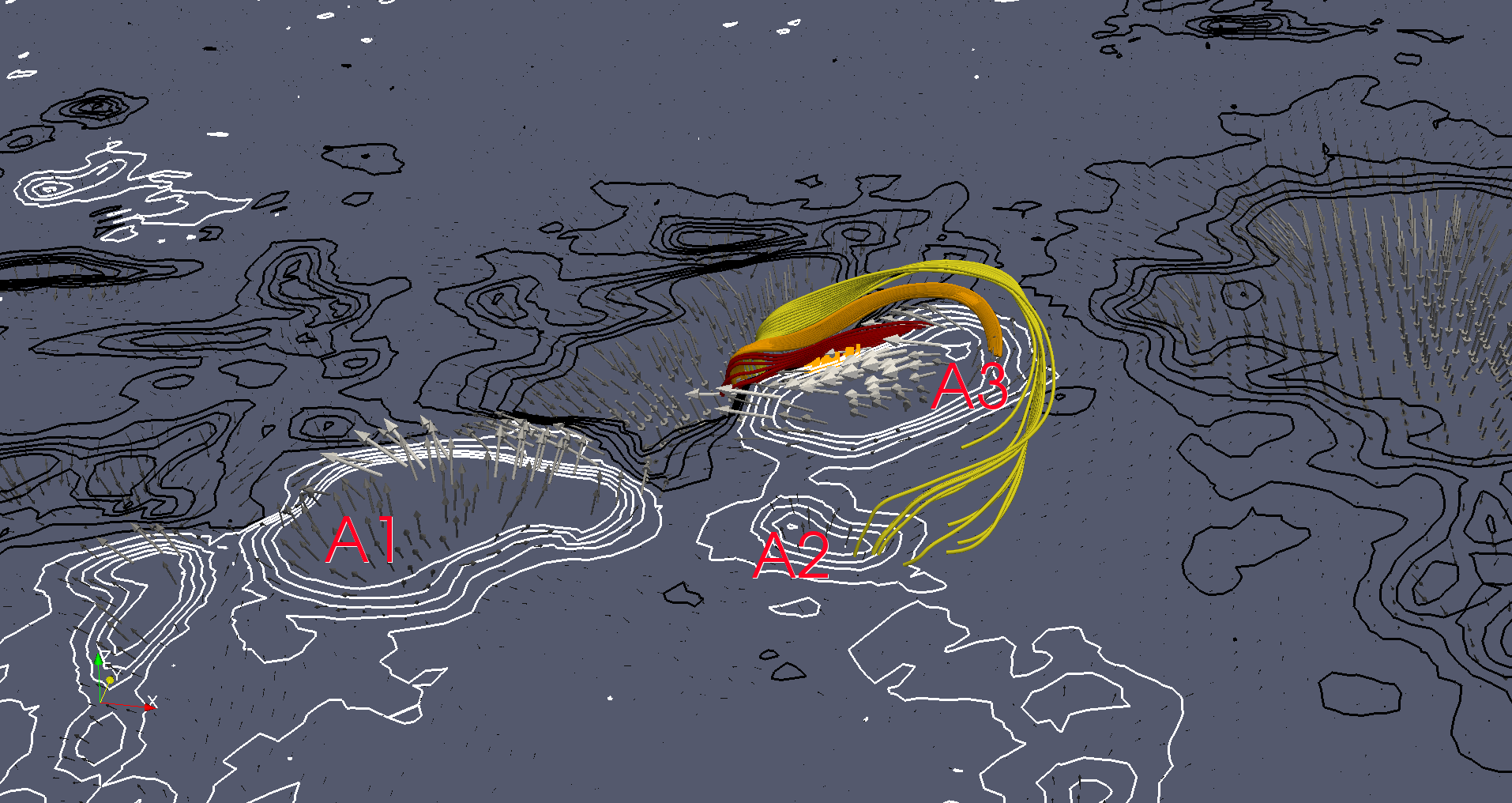}
\caption{\label{fig:BP}
Photospheric magnetic field in the vicinity of the positive field of Bipole A and negative field of Bipole B (\textbf{left}) with yellow squares marking the location of bald patches and arrows representing the transverse field, saturated at 1500~G. 
\textbf{Middle}: Three groups of field lines all starting along the vertical line located at the central bald patch. 
\textbf{Right} 3D-view of the same bald patch-related field lines, with arrows representing the magnetic field vector, saturated at 1500~G.
}
\end{center}
\end{figure*}

\subsection{The structure of the Quasi-Separatrix Layers at 14:36~UT}
\label{s:qsl}

\begin{figure*}
\setlength{\imsize}{0.48\TW}
\centering
\includegraphics[width=\textwidth]{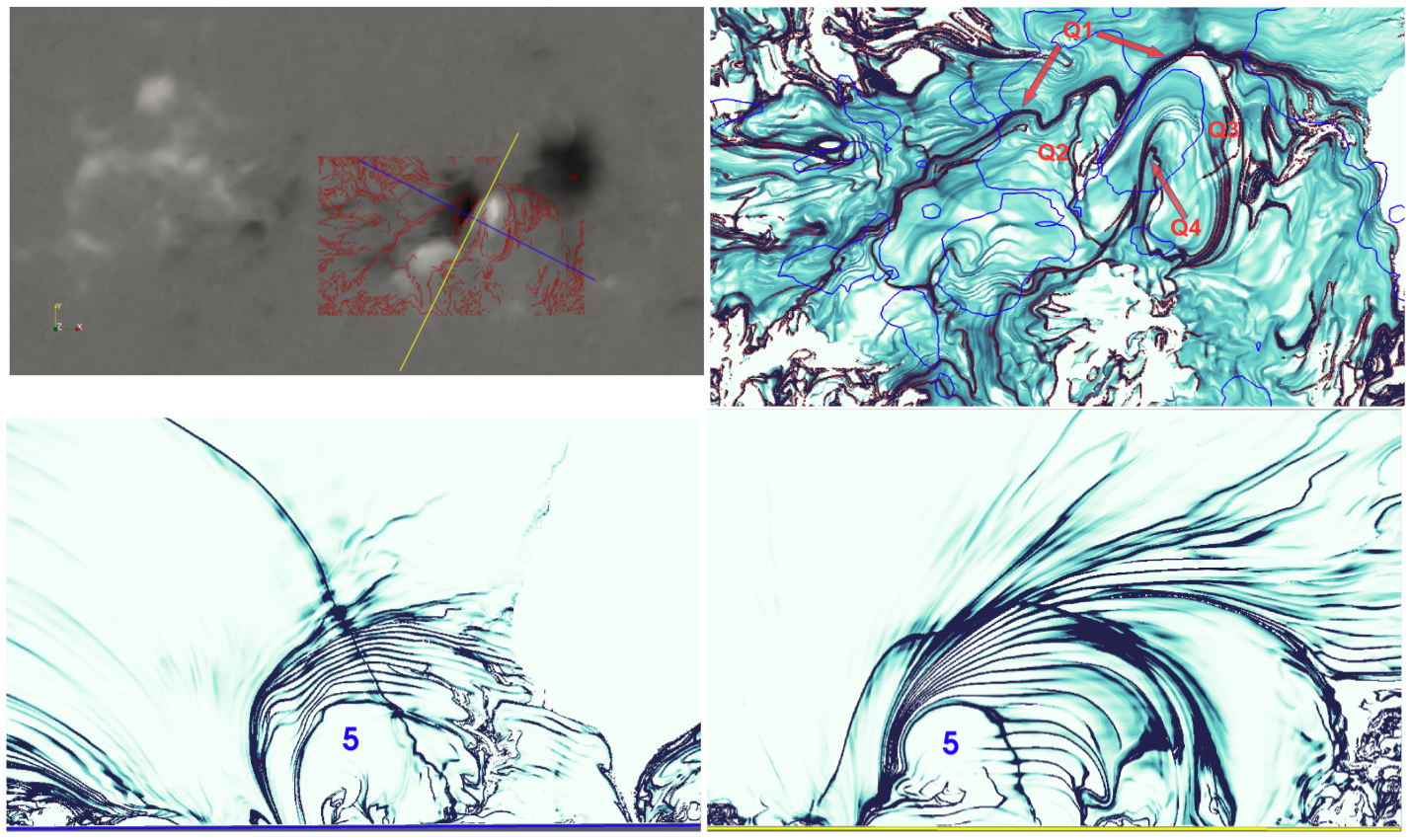}
\caption{\label{fig:mgm_qsl}Top Left: distribution of vertical magnetic field (grey-scale) at z=0.5 arcsec, overlaid with the Q-isoline (red) and the position of the vertical planes for the computation of Q distributions along (blue segment) and across (yellow segment) the flux rope's core; Top right: Q distribution (blue-scale) at z=0.5 arcsec, with the Q-isoline overlaid, and isolines of $B_{los}$ at 0 (blue line at 500G ). Bottom: Q distribution (blue-scale) on the vertical planes  across (left, over the blue segment in the top right panel) and along (right, over the green segment in the top right panel) the flux rope's core. Here and in all following plots: the red isoline of Q is plotted for Q=$10^{5}$; isocontours of Q in blue-scale cover values between 1 and $10^{6}$ in logarithmic scale, isocontours of $B_z$ cover values between -1500 G and 1500 G.
}
\end{figure*}

Quasi-separatrix layers \citep[QSLs,][]{Demoulinetal96} are regions where there is a drastic change in field line connectivity and are defined as locations where the squashing degree $Q$  is \textit{large} \citep{Tit2002}. Moreover, they are locations where perturbations can easily generate current sheets and magnetic reconnection can take place \citep{Aulanier2005}. Therefore, QSLs are of particular interest in the study of sunquake generation, which involves a release of energy from the coronal magnetic field and the propagation of energy and momentum to the lower atmosphere. In this section we compute the $Q$-factor using the latest version of the topology tracing code \citep[\textsc{topotr},][]{Demoulinetal96}, where the formula of \cite{Par2012} is implemented. The seed plane from which field lines are traced is at $z=0.5~$arcsec. Here we describe the different connectivity domains, as determined by the QSL computation, and the configuration of the magnetic field in these different domains. In the figures that follow, magnetic field lines are color-coded in the same way as the numbering of the connectivity domains. 

A 2D map of the computed QSLs along the horizontal plane at $z=0.5~$arcsec and covering the fan area is shown in the top-right panel of Figure~\ref{fig:mgm_qsl}. Two vertical Q-maps perpendicular to each other and crossing at (approximately) the bald patch locations, are shown in the bottom panels of Figure~\ref{fig:mgm_qsl}.
A rather complex connectivity structure is seen, with very sharp gradients (maximum Q values of the order of $10^{12}$). An important structure to note is the fan separatrix that is labelled Q1 in the top-right panel of Figure~\ref{fig:mgm_qsl}. As a comparison, the red field lines in Figure~\ref{fig:AIA_NLFFF_global_comparison} map to Q1, whereas the green field lines in the same figure are started just to the north of Q1. As we will see below, Q1 is indeed the location of the northern part of the circular ribbon observed during the C-class flare, as expected from other studies \citep[see \eg][]{Masson2009,Vemareddy2014,Yang2015,Masson2017, Zuc2017}. This is discussed further in Section \ref{s:discussion}.
The fan is also clearly visible in the outermost high-Q shell in the vertical $Q$-maps (bottom row panel in Figure~\ref{fig:mgm_qsl}). Note that the spine is \textit{not} included in those vertical sections (the vertical Q-line leaving the fan in the north-east direction in Figure~\ref{fig:mgm_qsl} lower-left panel is related to a small, low-lying null outside the fan roughly corresponding to the indent in Q1). 

Under the fan, a layered structure is found that corresponds to field line compression and rarefaction created by the granularity of one of the polarities: field lines originating from a strong compact polarity (negative in this case) are connected to a weaker, more fragmented polarity, like the polarity A1 and the elongated boundary of the super-granular cell to the south of it. As a result, field lines cluster in two-dimensional bundles.
A similar structure was found by \cite{Masson2017}, see in particular their Fig.~5. 
Below the layered structure, a uniform, low Q-value area is present, labeled 5 in the bottom row panels in Figure~\ref{fig:mgm_qsl}. However, below domain 5 there is a fine structured high-Q concentration region close to the photosphere. A zoom in of this area is presented in Figure~\ref{fig:HFT}, where additional labels mark the main connectivity domains (we ignore here the domain surrounded by 2,3, and 5 since it shows a similar connectivity to 3, as well as smaller domains). 

A summary of all the connectivity domains below the fan, with their associated field lines, is shown in Figure~\ref{fig:fl_below_fan}. We now discuss the importance of these different sets of field lines in domains 1 (red field lines), 2 (orange field lines), 3 (light-yellow field lines), 4 (blue field lines) and 5 (green field lines) in the context of the overall configuration. 

\setlength{\imsize}{\CW}
\begin{figure}
\begin{center}
\includegraphics[width=\imsize]{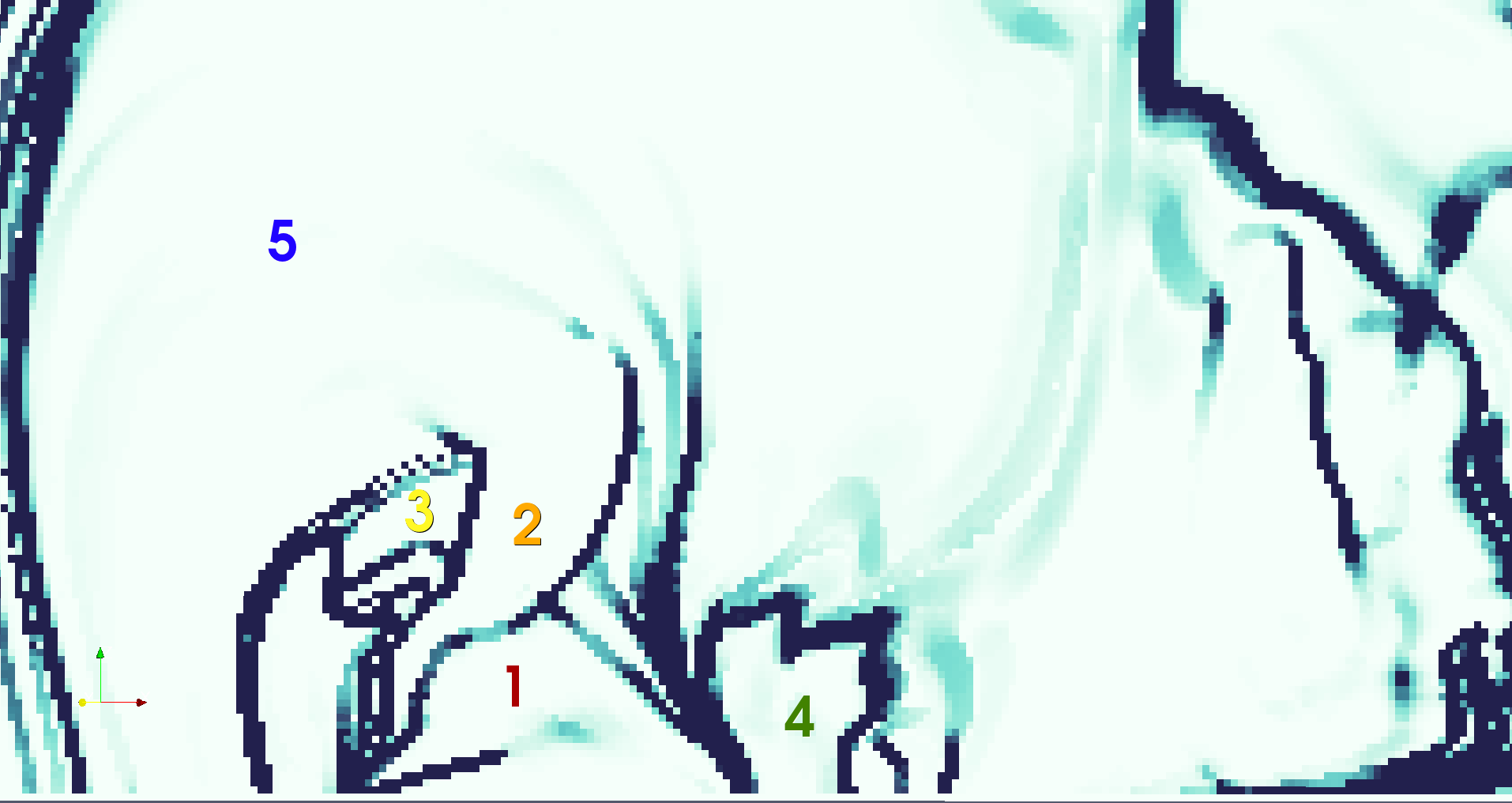}
\includegraphics[width=\imsize,clip,trim= 400 0 0 0]{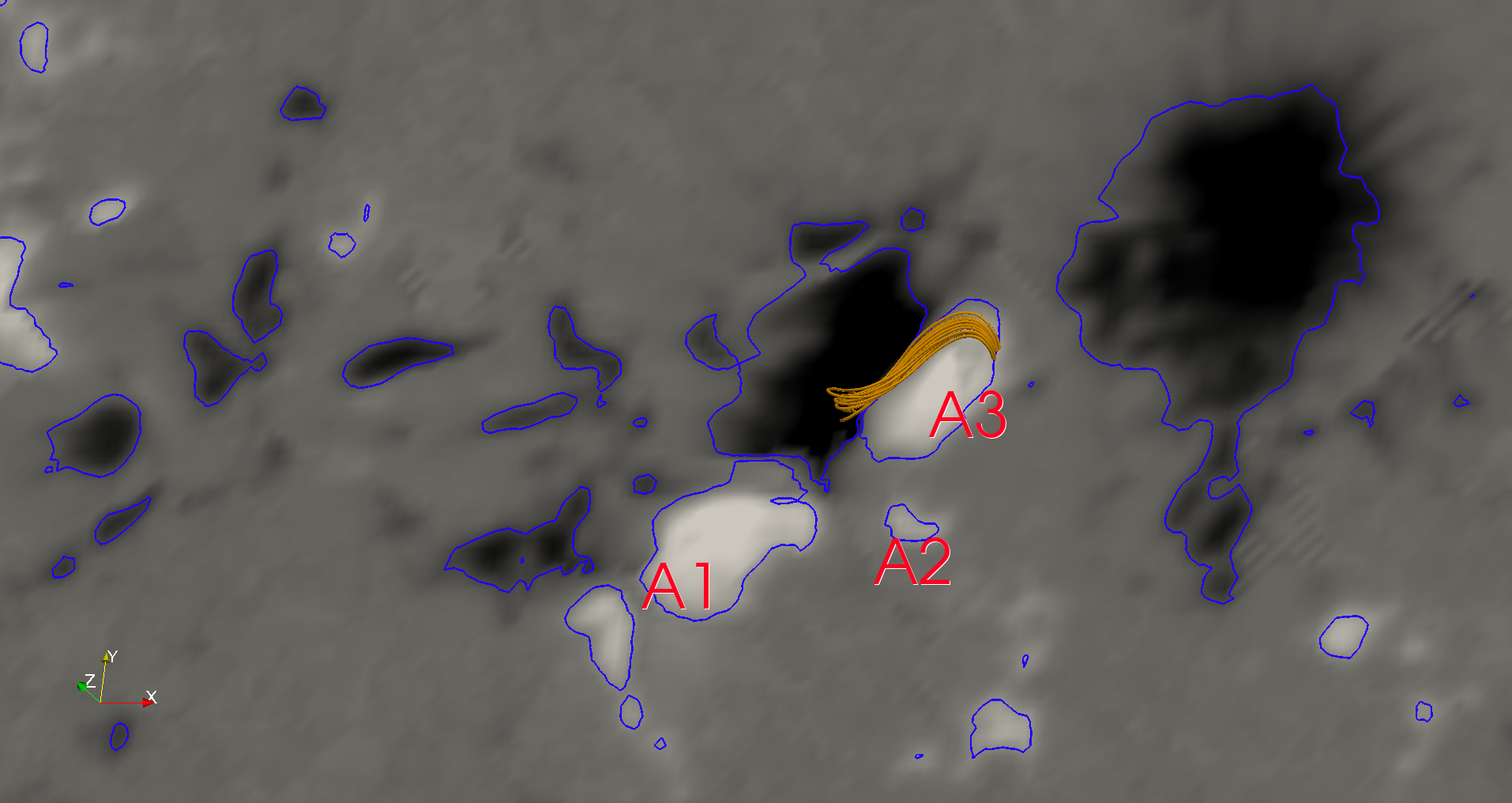}
\end{center}
\caption{\label{fig:HFT}
\textbf{Top}: Zoom view of the vertical Q-map  in the bottom left panel of Figure~\ref{fig:mgm_qsl} showing the different domains of connectivity. Colored labels correspond to field lines in all plots. 
\textbf{Bottom} Field lines started from the connectivity domain 2, right above the HFT. These field lines compare favorably with the filament of Figure~\ref{fig:AIA_335_HMI_1538} especially accounting for the additional northward migration of the polarity A3 in the 70 minutes time difference between the two figures.
}
\end{figure}


\begin{figure*}
\centering
\includegraphics[width=\textwidth] {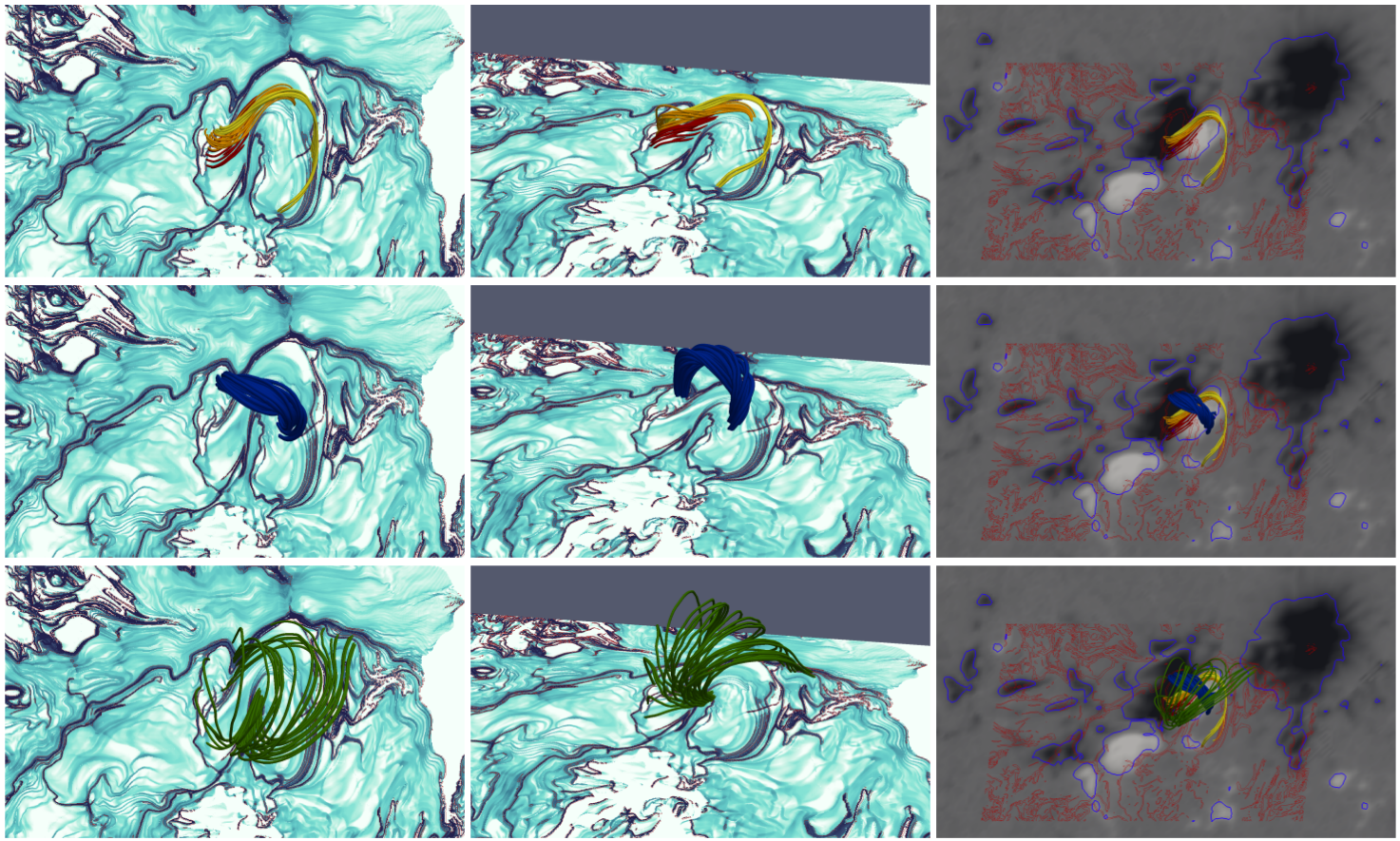}
\caption{\label{fig:fl_below_fan} 
Field lines corresponding to the main connectivity domains below the fan. 
From top to bottom: \textbf{First row}: field lines traced from below the HFT (red), right above the HFT (orange), above the HFT sheared (light-yellow). 
\textbf{Second row}: Bulk of the flux rope.
\textbf{Third row}: Sheared envelope field lines (green)
Top (\textbf{first column}) and 3D view (\textbf{middle column}) on photospheric Q-distribution (blue-scale) and Q isoline (red), and observer view (\textbf{right column}) on magnetogram (grey-scale) with overlaid Q isoline. 
}
\end{figure*}

Let us first consider the three groups of field lines passing through domains 1, 2 and 3. The three groups of field lines all start from a narrow, high-Q concentration in the negative polarity of Bipole B, between QSL Q2 and the polarity inversion line (cf. Figure~\ref{fig:mgm_qsl} and the top row of Figure~\ref{fig:fl_below_fan}).
The red field lines starting from domain~1 are strongly sheared field lines, largely parallel to the polarity inversion line, and connect the inner sides of the polarities A3 and the negative polarity of Bipole B. Above them are field lines starting from domain~2 that are sheared and have a sigmoidal shape. These field lines are shown in orange. They have clear dips and connect the negative polarity of Bipole B to the polarity A3. 
The third group of field lines, starting from domain 3 and rendered in light-yellow circle around polarity A3 and connect to the more distance polarity A2 that formed from the fragmentation of the positive polarity of Bipole A.
The high-Q values between domains 1 and 2 in Figure~\ref{fig:HFT} supports the interpretation of the presence of an HFT in this location which separates the red and orange field lines in Figure~\ref{fig:fl_below_fan}.

It is noted that the field lines just above the HFT (orange field lines in Figure~\ref{fig:HFT} and Figure~\ref{fig:fl_below_fan}) have a very similar shape and location to the filament shown in the top panel of Figure~\ref{fig:AIA_335_HMI_1538}. The filament is most clearly visible shortly before the eruption rather than at the time of the extrapolation 70 minutes earlier, \ie after the additional northward migration of the polarity A3 has taken place. Such a migration indeed improves the qualitative match between the observed filament and the orange field lines in the extrapolation by elongating the orange structure in the north direction, making the association very convincing. This is more evident by comparing the relative location of the photospheric anchoring of the orange field lines in  Figure~\ref{fig:HFT} with the filament ends marked by the red plus signs on Figure~\ref{fig:AIA_335_HMI_1538}. Therefore, the filament is most likely associated with the field lines right above the HFT, which are alinged with the polarity inversion line at their core, hooked at both ends, and run low in the atmosphere. However, given the difference in time, we refrain from overlay the corresponding images.

Domain~2 (the orange filament field lines) is actually not separated from domain~5 by any QSL, that is, there is a smooth transition between field lines of the two domains. However, field lines starting from the core of domain~5 (Figure~\ref{fig:HFT} and blue field lines in Figure~\ref{fig:fl_below_fan}) show a right-handed, moderately twisted flux rope core that is actually almost \textit{perpendicular} to the polarity inversion line. Therefore, we keep the two domains visually separated to better show this unusual configuration. The bottom part of the flux rope is separated by an HFT from the photosphere (as is usual when the flux rope's axis is high enough in the corona) while field lines below the HFT are essentially aligned with the polarity inversion line. The bottom part of the flux rope is at angle with the polarity inversion line in its center and wraps around the flux rope feet at its ends (in a right-handed sense), but the angle is rather large. Such an angle increases further for field lines closer to the flux rope axis until it is almost 90 degrees for the core itself. The feet of the flux rope are surrounded by two high $Q$-values hooks, marked as Q1 and Q3 in the top-right panel of Figure~\ref{fig:mgm_qsl}.

Finally, field lines starting from domain~4 form an envelope surrounding the flux rope and they are essentially perpendicular to its central section (see the green field lines in Figure~\ref{fig:fl_below_fan}). Note that such field lines originate from a very narrow domain (marked as Q4 in Figure~\ref{fig:mgm_qsl}), surround the flux rope core, and cross the vertical planes at the round QSL bounding region 5 upwards. Such field lines are also at a finite angle to the field lines right above them, \ie belonging to the layered domain, some example of which are the cyan field lines in Figure~\ref{fig:AIA_NLFFF_core_comparison}.

\section{Scenario of the active region evolution leading to the sunquake}
\label{s:discussion}

Combining the multi-wavelength observations discussed in Section \ref{s:observations} with the snapshot of the coronal magnetic field from the NLFFF model analyzed in Section \ref{s:nlfff}, we are now able to sketch a scenario for the evolution leading to the C-class and M-class flares, the filament eruption and to the associated sunquake. 

At the larger scales, the configuration of the magnetic field has the shape of a fan and spine generated by a coronal null point - a key aspect of the configuration where magnetic reconnection can occur if the null point is perturbed. The part of the active region that exhibits dynamic activity involves mostly the fan, and the structures beneath it.


At the photopsheric level, the polarity inversion line between  bipoles A and B of Figure~\ref{fig:LOS_evolution} shows the presence of bald patches under the location where the filament is later observed. The bald patches remain until 15:48 UT, which is the time of the peak of the C-class flare and one minute before the onset of the M-class flare.
At the time of the NLFFF snapshot (17 February 14:36~UT), the magnetic field has the configuration of a HFT surrounded by a high current concentration.
In such a configuration, plasma heating is expected to occur on field lines that run through the HFT due to Ohmic dissipation, creating the hot plasma emission structures that are seen running close to the PIL in the AIA data.
In the extrapolation, such field lines indeed carry strong electric currents as is expected to happen at such sensitive locations, see \eg \cite{Aulanier2005}.
Given the persistent presence of bald patches for several hours before the eruption, and the HFT configuration found by the extrapolation 70 minutes before it, it is likely that the collision of bipoles A and B, and the shearing along the polarity inversion line between them, is responsible for the filament formation observed shortly before the eruption. 


Above the HFT there are the flux rope field lines. 
We know from the above discussion and the extrapolation that the  flux rope is formed at least one hour before eruption, likely by reconnection at the HFT.
The bald patch-HFT system is responsible for the creation of the surrounding helical structure, and likely for its destabilization, confirming the HFT's role in eruptions as in \cite{Masson2017}. Similarly to that paper, the current distribution at the HFT demonstrates that reconnection can occur there (and reproduce pre-eruptive observations). 



The splitting of the positive polarity of Bipole A drags the field lines and creates new, non-standard signatures in the Q-map. While the polarity A1 stays unchanged, relatively un-sheared and with no activity, polarity A3 moves northwards, shearing the field and creating the bald patches and the bald patch-to-HFT transition, conceptually similar to the evolution discussed by \cite{aulanier2010}. Polarity A2, which was connected to the bottom part of A3 and which is moved sideways by its northern migration, carries with it part of the field lines, which become our light-yellow field lines of Figure~\ref{fig:fl_below_fan}. Conversely, the geometrical arrangement of the light-yellow connectivity and the above interpretation of the photopsheric evolution seem to imply that the flux bundle that becomes the blue flux rope was already present before the separation of the polarity A started, complementing in this way the picture coming from the observations alone.

It is interesting to note that, as consequence of such a complex photospheric dynamics, the flux rope core rotates until it is aligned practically perpendicular to the PIL, to the point that the filament field lines (identified with field lines right above the HFT) and the flux rope are almost perpendicular to each other. We explain this unusual orientation by the strong northward motion of the positive polarity A3, which moves the footpoints away from their earlier position to the south, which would then have been more spatially correlated with the axial field of the flux rope. 

\begin{figure}
\centering
\includegraphics[width=\columnwidth]{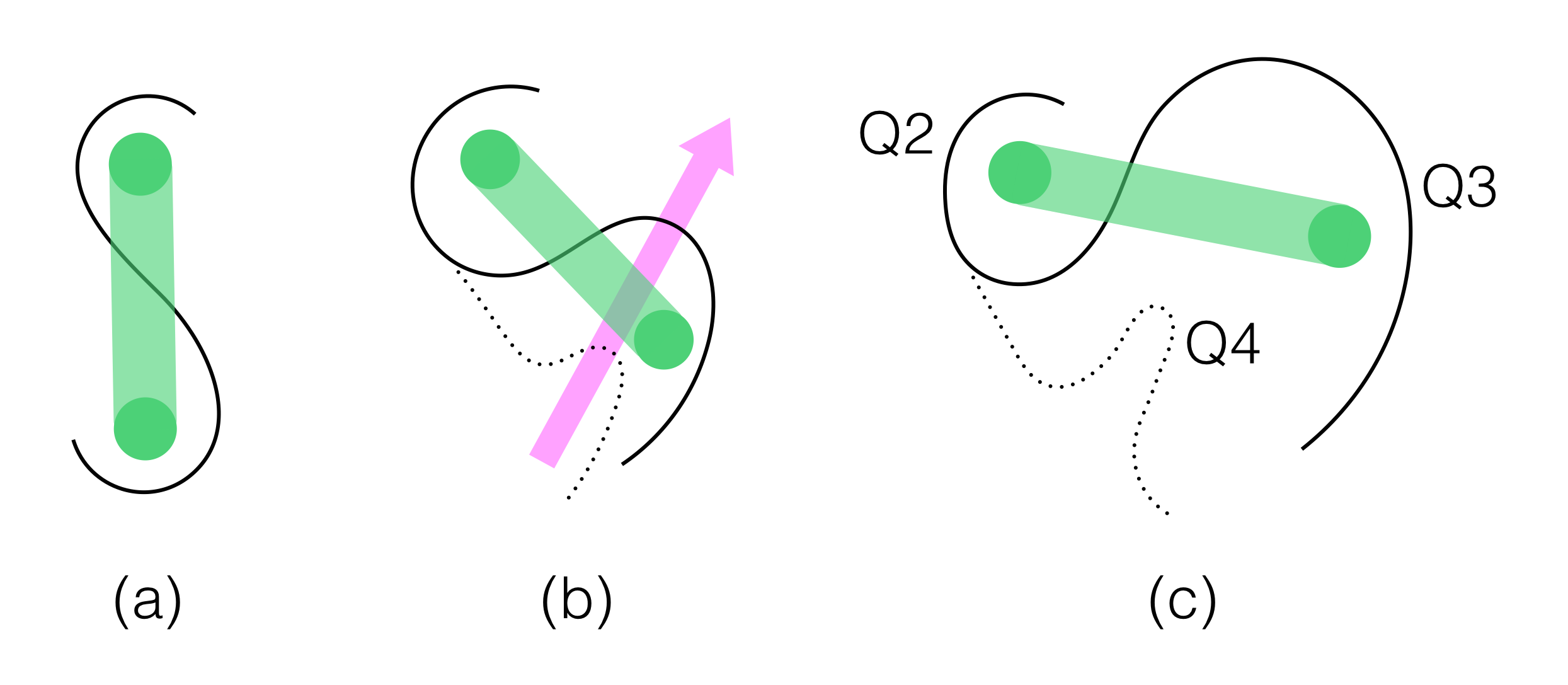}
\caption{\label{fig:qsl_shearing}
Schematic showing how a standard configuration flux-rope (shown in  green) and QSL (black line, panel a) can be modified by the observed shearing motions. The splitting of the polarity and its northward motion create a new branch in the QSL (dashed line in panels b and c) and rotate the foot of the flux rope until a flux rope configuration and a QSL distribution similar to the one reconstructed by the NLFFF extrapolation is obtained. The magenta arrow illustrates the northwest motion of the positive polarity A3.
} 
\end{figure}

Figure~\ref{fig:qsl_shearing} shows a schematic illustrating the effect of such a process on the QSL and flux rope locations: the leftmost panel shows the ``standard'' configuration of a flux rope and its associated main QSL (without interruption below the flux rope, hence with bald patches), as shown for instance in Figure~4a of \cite{Titov1999}. The split of the southern polarity, and the northward migration of the northern-most polarity has three effects: first, it drags the southern footpoint of the flux rope northward; second, it deforms the main QSL by rotating its southern hook anti-clockwise; third, it generates a subsidiary branch of the QSL related to the field lines rooted in the two split polarities (middle panel in Figure~\ref{fig:qsl_shearing}). The northward migration proceeds until the observed distribution is obtained, yielding a configuration that is modelled by the extrapolated field, namely a flux rope whose axis is essentially perpendicular to both the filament and the PIL below it, and a peculiar QSL distribution (rightmost panel in Figure~\ref{fig:qsl_shearing}).  

Therefore, some of the complexity of the Q-map in Figure~\ref{fig:mgm_qsl} is the result of a geometrical deformation essentially due to the northward migration of the polarity A3.
With that in mind, we can recognize the more standard J-shaped high-Q signatures of the flux rope in the Q2 (around the eastern end of the flux rope) and Q3 (around its western end), with the usual location of the bald patches. 
Note that Q2 and Q3 are practically merging into Q1 (the QSL in the northern side of the the fan), \ie the structures inside the fan that generate the Q2 and Q3 QSLs are indeed very susceptible to an interaction with the fan itself whenever any one of them is perturbed.
An additional product of the splitting of the positive polarity of Bipole A, and the northward migration of A3, is the creation of a secondary branch in the Q-maps between the J-shaped concentrations resulting in the formation of the structure Q4, origin of the very focused envelope field lines surrounding the flux rope (green field lines in Figure~\ref{fig:fl_below_fan}).



\begin{figure}
\centering
\includegraphics[width=0.5\textwidth]{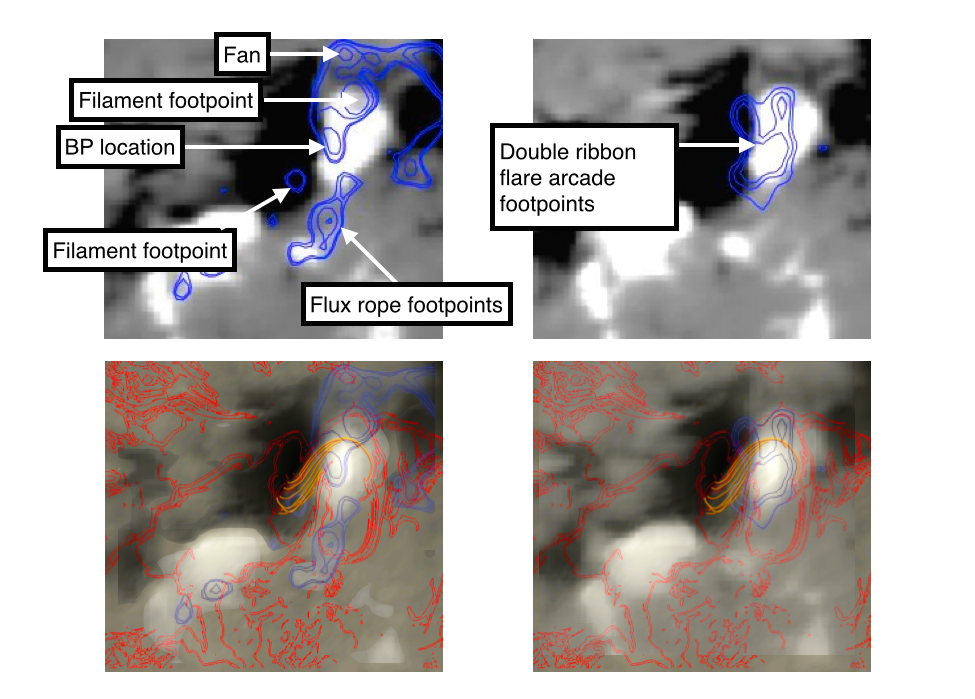}
\caption{\label{fig:qsl_flare} Qualitative relation between the Q map (red countours in bottom row) and the flare ribbons (blue contours) for the C-class (left) and M-class (right) flares. Tentative matching should account for the 70 minutes difference between the Q isolines from the extrapolation and the flare ribbons observations.
} 
\end{figure}

The northward migration of the positive polarity A3 (in which the filament is rooted) has foreseeable consequences for the stability of the flux system. As demonstrated in numerical simulations by, \eg \cite{Toeroek2013} and \cite{Zuccarello2012}, increasing the writhe/shear of the field connected to the flux rope region builds free magnetic energy into the system. In turn, the increase in energy leads to an expansion of the system, and its possible destabilization, on the same time scale as the trigger, \ie the northward migration of the polarity A3 in our case.

This sunquake study includes the analysis of two flares that occur sequentially, very close in time, and in approximately the same location; a C-class flare and an M-class flare. These two flares, however, have differently shaped flare ribbons (Figure~\ref{fig:1700_ribbons}), which shows that they are distinct energy release events.
In the first flare (C-class flare), the ribbon has a circular form, which is associated with a null point activation suggesting that reconnection took place at the null. The NLFFF extrapolation confirms that the QSL of the fan is at the same location as the circular ribbon (see Figure \ref{fig:qsl_flare}). To the south of the circular ribbon there are other brightenings that correspond spatially with the bald patch locations and the footpoints of field lines above (and possibly also below) the HFT. We propose that the C-class flare is the result of a first phase of the eruption involving reconnection at the null, likely due to the expansion of the flux system below it. Such an expansion, by the usual flare reconnection, also has an associated reconnection at lower altitudes, as observed. 

The second flare is the M-class flare and it is associated with the ejection of the filament material. The flare is generated by structures that are internal to the fan (cf Figure~\ref{fig:qsl_flare}) and exhibits ribbons that are similar in shape to a two-ribbon flare. The location of the flare ribbons can be associated with the QSLs close to the polarity inversion line between Bipole B and A3 and the QSL on the western side of A3. The flare ribbons are elongated along the QSLs, and do not seem to involve the fan QSL Q1 at all. On these grounds, we argue that the second flare is related to the eruption of the flux rope itself, after which the filament material is clearly being expelled (figure~\ref{fig:aia_images}) and the bald patches disappear. Hence, the M-class flare is a consequence of the proper ejection of the flux rope evidenced by the formation of double flare ribbons.

The time of the sunquake 
and its location suggest its connection with field lines rooted in the northward migrating polarity A3. 
Given the difference in time between the extrapolation and the sunquake, we are not able to directly connect its location with a specific set of field lines, an operation that is further complicated by the uncertainties in the alignment discussed in Section \ref{s:aia_comparison}. However, taking into account the discussion so far, our best guess associates the sunquake location (taken to be the 5~mHz signal in Figure~\ref{fig:egression_ic_mg}) to the connectivity domain indicated as Q4 in Figure~\ref{fig:mgm_qsl}. That connectivity domain corresponds to the domain~4 in Figure~\ref{fig:HFT}, \ie to the envelope bundle of field lines  (in green in Figure~\ref{fig:fl_below_fan}). Given its structure, such a field line bundle has two interesting properties. First, since it is  surrounding the flux rope in almost its entire length it is then directly impacted when the eruption occurs.
Second, such wide-spread field lines between the erupting flux rope and the above layered structure are all connected on one side to a very small area in the photosphere (where the sunquake occurs), in this way providing a sort of magnetic focusing.
In light of this, we argue that the sunquake is a consequence of the energy release associated to the reconnection between the erupting flux rope and the envelope surrounding it. Furthermore, the envelope structure helps focus the released energy into a small area in the lower atmosphere (Q4 in Figure~\ref{fig:mgm_qsl}) which produces the sunquake pulse.

The scenario offered by our analysis and the interpretation of the observations clarifies the dynamics of the two flares and supports the association of the sunquake to the larger, M-class flare, rather than to the first C-class flare. 
Moreover, it indicates how a sunquake may require special magnetic field configuration to occur, which may help understand their erratic occurrence.


\section{Conclusion}
\label{s:conclusion}

In this work we study the evolution of active region 11675 on 17 February 2013, in the hours leading to the production of two flares (a C-class flare and an M-class flare), the ejection of filament material and a sunquake. We investigate the configuration of the active region magnetic field through the construction of a non-linear force-free field (NLFFF) extrapolation just over one hour before the sunquake. The aim of the work is to shed light on the configuration of the coronal magnetic field in which the sunquake was produced.

The sunquake occurs when active region 11675 is undergoing a period of rapid evolution during its emergence phase, and when there are strong motions of the photospheric magnetic field elements. In particular, two emerging bipoles collide and shear past each other. The magnetic field along the polarity inversion line that separates these bipoles is highly sheared and at low altitudes bald patch field lines are present. At higher altitudes a hyperbolic flux tube (HFT) and a flux rope are present, which have been highly modified by the motions of the photospheric field in which these magnetic structures are rooted. 

From a time-distance analysis we determine that the sunquake was initiated in the time period between 15:50:41 to 15:51:26 UT on 17 February 2013 in a region of positive polarity field. This finding indicates that the sunquake is initiated during the M-class flare, which is associated with the eruption of the flux rope and ejection of filament material. However, this timing determination differs from the result presented in \cite{Sharykinetal15}, who suggest the sunquake onset time to be 15:47:54 UT and therefore propose that the sunquake is associated with the earlier C-class flare. The difference in sunquake initiation time in these two studies arises from the different techniques used, the different source position and the interpretation of a transient velocity signal in \cite{Sharykinetal15}. Here, we combine acoustic holography and time-distance analysis techniques whereas \cite{Sharykinetal15} use an observed velocity transient that occurred at 15:47:54 UT. Our source location represents the location that yields a sharper, and hence more accurate, ridge reconstruction in the time-distance analysis. The question of photospheric velocity transients and sunquake association is an interesting one, and will be the subject of future research.


The $Q$ maps derived from the NLFFF extrapolation are the first steps for an investigation of the site of energy release and the transfer of energy and momentum from the corona to the lower atmosphere that could trigger the sunquake. The extrapolation can be used to probe
the coronal configuration to understand why it was the M-class flare, and not the C-class flare that showed a more impulsive hard X-ray light curve than the M-class, that produced the sunquake. Given the spatial correspondence of the sunquake's origin and the location of $Q$4, we propose that the reconnection site responsible for the energy release that is ultimately  related to the sunquake is all around the top of the flux rope, and that this reconnection, triggered by the expansion of the erupting flux rope, liberates energy along the $Q$4 field lines (green field lines in Figure \ref{fig:fl_below_fan}). This scenario corroborates  the association of the sunquake to the M-class flare, since the reconnection at the top would start at the same time as the flux rope instability, which in turn is the same time as the reconnection under the flux rope that is responsible for the M-class flare and its associated double-ribbon.

In conclusion, our investigation supports the idea that in this event the sunquake's occurrence and location are strongly influenced by the magnetic field configuration and its dynamic evolution. A configuration is present that may act as a magnetic "lens", focussing the released energy to a particular site in the lower atmosphere. 
The scope of this study does not permit us to provide information on the physical mechanisms that generated the sunquake in this event. Instead, our approach is to investigate in unprecedented detail in the local magnetic environment in which the sunquake occurs. However, we do note that the magnetic field configuration is relevant to both wave and particle sunquake mechanisms, as the magnetic field acts as a guide for both waves and particles. Such a particular magnetic lens configuration, as found in active region 11675, may not always be present in solar flare magnetic fields, which could explain the erratic nature of sunquake occurrence.



\acknowledgments
LMG and GV acknowledge the support of the Leverhulme Trust Research Project Grant 2014-051. LMG also thanks
the Royal Society for funding through their University Research Fellowship scheme. F.P.Z. is a Fonds Wetenschappelijk Onderzoek (FWO) research fellow (Project No. 1272714N). SAM acknowledges funding from the STFC under the Consolidated Grant ST/N000722/1. SLG acknowledges funding from the Italian MIUR-PRIN grant 2012P2HRCR on the active Sun and its effects on space and Earth climate, by the Space Weather Italian COmmunity (SWICO) Research Program.

\bibliographystyle{plainnat}
\bibliography{references}

\end{document}